\PassOptionsToPackage{dvipsnames}{xcolor}
\documentclass{aastex631}
\pdfoutput=1
\usepackage{amsmath}
\usepackage{capt-of}   % don't use the package caption -- it causes some kind of conflict.  7/29/15
\usepackage{empheq}
\usepackage{enumitem}
\usepackage{hyperref}
\usepackage{hhline}
\usepackage{numprint}
\usepackage{pifont}
\usepackage{xspace}
\usepackage{placeins}
\usepackage{graphicx}
\usepackage{mwe}
\usepackage{multirow}
\usepackage{tabularx,booktabs}
\usepackage{hhline}
\usepackage{booktabs}
\usepackage{changepage}
\usepackage{amsmath}
\usepackage{slashbox}

\newcommand{\grandtotnew}{4,869\xspace}%total accross all papers including this work

\newcommand{\NnewlensesUndisc}{811\xspace}
\newcommand{\NnewlensesUndiscA}{90\xspace}
\newcommand{\NnewlensesUndiscB}{104\xspace}
\newcommand{\NnewlensesUndiscC}{617\xspace}

\newcommand{\Ninspectedsystems}{5680\xspace}

\newcommand{\Nnewlensesacrosspapers}{3868\xspace}

\newcommand{\SERpurity}{5\xspace}
\newcommand{\DEVpurity}{18\xspace}
\newcommand{\REXpurity}{18\xspace}
\newcommand{\EXPpurity}{80\xspace}

\newcommand{\ho}{\ensuremath{H_0}\xspace}

\newcommand{\tractor}{\textit{The Tractor}\xspace}

\newcommand{\twopr}{^{\prime \prime}}

\newcolumntype{Y}{>{\centering\arraybackslash}X}

\shorttitle{Strong Lenses in DESI Legacy Surveys DR10}
\shortauthors{Inchausti, Storfer, Huang et al.}

\begin{document}
\title{Strong Lens Discoveries in DESI Legacy Imaging Surveys DR10 \\
with Two Deep Learning Architectures}

\correspondingauthor{Jose Carlos Inchausti, Xiaosheng Huang}
\email{jinchaustireyes@usfca.edu, xhuang22@usfca.edu}

\author[0009-0009-8667-763X]{Jose Carlos Inchausti}
\affiliation{Department of Physics \& Astronomy, University of San Francisco, San Francisco, CA 94117, USA}

\author[0000-0002-0385-0014]{Christopher J.~Storfer}
\affiliation{Institute for Astronomy, University of Hawai`i, 2680 Woodlawn Drive, Honolulu, HI 96822, USA}
\affiliation{Physics Division, Lawrence Berkeley National Laboratory, 1 Cyclotron Road, Berkeley, CA, 94720}

\author[0000-0001-8156-0330]{Xiaosheng~Huang}
\affiliation{Department of Physics \& Astronomy, University of San Francisco, San Francisco, CA 94117-1080}
\affiliation{Physics Division, Lawrence Berkeley National Laboratory, 1 Cyclotron Road, Berkeley, CA, 94720}

\author[0000-0002-6876-8492]{Yuan-Ming~Hsu}
\affiliation{Department of Physics, National Taiwan University, No. 1, Section 4, Roosevelt Road, Taipei 106319, Taiwan}

\author[0009-0002-4310-3539]{Brandt Kaufmann}
\affiliation{Department of Mathematics \& Statistics, University of San Francisco, San Francisco, CA 94117, USA}

\author{Chaitanya Pasupala}
\affiliation{Department of Economics, University of San Francisco, San Francisco, CA 94117, USA}

\author{S.~Banka}
\affiliation{Department of Electrical Engineering \& Computer Sciences, University of California, Berkeley, Berkeley, CA 94720}

\author[0000-0002-4928-4003]{A.~Dey}
\affiliation{NSF's National Optical-Infrared Astronomy Research Laboratory, 950 N. Cherry Ave., Tucson, AZ 85719}

\author[0000-0002-1172-0754]{D.~Lang}
\affiliation{Perimeter Institute for Theoretical Physics, Waterloo, ON N2L 2Y5, Canada}

\author[0000-0002-1125-7384]{A.~Meisner}
\affiliation{NSF's National Optical-Infrared Astronomy Research Laboratory, 950 N. Cherry Ave., Tucson, AZ 85719}

\author[0000-0002-2733-4559]{J.~Moustakas}
\affiliation{Department of Physics and Astronomy, Siena College, 515 Loudon Rd., Loudonville, NY 12211}

\author{A.D.~Myers}
\affiliation{Department of Physics \& Astronomy, University of Wyoming, 1000 E. University, Dept 3905, Laramie, WY 82071}

\author[0000-0002-3569-7421]{E.~F.~Schlafly}
\affiliation{Space Telescope Science Institute, 3700 San Martin Drive, Baltimore, MD 21218, USA}

\author[0000-0002-5042-5088]{D.J.~Schlegel}
\affiliation{Physics Division, Lawrence Berkeley National Laboratory, 1 Cyclotron Road, Berkeley, CA, 94720}

\begin{abstract}
We have conducted a search for strong gravitational lensing systems in the Dark Energy Spectroscopic Instrument (DESI) Legacy Imaging Surveys Data Release 10 (DR10). This paper is the fourth in a series of searches \citep[following][Paper~I, II, \& III respectively]{huang2020a, huang2021a, storfer2024a}. This is the first catalog of lens candidates covering nearly the entirety of the extragalactic sky south of declination $\delta~\approx +32^{\circ}$, all of it observed by the DECam, covering $\sim$14,000 $deg^2$. We impose a $z$-band magnitude cut of $<$ 20 in AB magnitude. We deploy a Residual Neural Network and EfficientNet as an ensemble trained on a compilation of known lensing systems and high-grade candidates as well as nonlenses in the same footprint. The predictions from these two base models are aggregated using a meta-learner. After applying our ensemble to the survey data, we exclude known candidates and systems, and use our own visual inspection portal to rank images in the top 0.01 percentile of all neural network recommendations. We have found 811 new lens candidates. These include 484 new candidates in the Legacy Surveys DR9 footprint, all parts of which have been searched for strong lenses at least once before, either by our group or others. Combining the discoveries from this work with those from Paper I (335), II (1210), and III (1512), we have discovered a total of \Nnewlensesacrosspapers \textit{new} candidates in the DESI Legacy Surveys.
\end{abstract}
% \vspace{0.5in}
\section{Introduction}
\label{sec:intro}
Strong gravitational lensing systems are a powerful tool for cosmology. They have been used to study dark matter in galaxies and clusters \citep[e.g.,][]{kochanek1991a, broadhurst2000a, koopmans2002a, bolton2006a, koopmans2006a, massey2010a, grillo2015a, shu2017a},  
and are uniquely suited to probe substructure in cluster and galaxy scale lenses, 
as well as line-of-sight low-mass halos and test the predictions of the cold dark matter (CDM) model beyond the local universe \citep[e.g.,][]{vegetti2009a,  huang2009a, vegetti2010a, meneghetti2020a, meneghetti2023a,  cagansengul2021a, wagner-carena2022, fagin2024a}.
Searches and spectroscopic follow-up of lensed quasars \citep[e.g.,][]{dawes2023,sheu2024,he2023} have enabled time-delay cosmography, offering an independent route to measure \ho.
Multiply lensed supernovae (SNe) are also ideal for measuring time delays and \ho because of their well-characterized light curves, and in the case of Type~Ia, 
with the added benefit of standardizable luminosity \citep[e.g.,][]{refsdal1964a, treu2010a, oguri2010a}, provided microlensing can be accurately characterized \citep{yahalomi2017a}.
Furthermore, lens models can be constructed after the SNe have faded \citep{ding2021a}.
In the last decade, strongly lensed supernovae,  both core-collapse \citep{rodney2016a,chen2022a} and Type~Ia \citep[e.g.,][]{pierel2023a,pierel2024a,pascale2024a} have been discovered. \citet{sheu2023a} conducted a retrospective search for strongly lensed supernovae in the Dark Energy Spectroscopic Instrument (DESI) Legacy Imaging Surveys \citep{Dey2019}, and found seven promising candidates.
Time-delay \ho measurements from multiply imaged supernovae \citep[e.g.,][]{pierel2019a, suyu2020a, huber2021a,Pascale_2025}, 
combined with measurements from distance ladders \citep[e.g.,][]{freedman2020a,riess2021a} and lensed quasars \citep[e.g.,][]{wong2019a,millon2020a,birrer2020a}, can be an important test of the tension  between \ho measured locally and the value inferred from the Cosmic Microwave Background \citep[CMB;][]{planck2020}.
In addition, magnified (but not multiply-lensed) SNe~Ia were identified \citep{patel2014a, nordin2014a, rubin2018a} and used to test the lens models.
Finally strong lensing systems can be used to constrain the properties of dark energy \citep[e.g.,][]{jullo2010a, li2024a}, especially using systems with multiple source planes \citep[e.g.,][]{collett2014a, linder2016,  sharma2022a, caminha2022a}.

The introduction of neural networks to identify gravitational lens candidates in imaging surveys has been transformative and continues to show success in identifying lensing candidates \citep[e.g.,][]{jacobs2017a, metcalf2018a, jacobs2019a, jacobs2019b, canameras2020a,canameras2021a,rojas2022a,shu2022a,Zaborowski_2023,gonzales2025}. 
In our recent work, we discovered over 3000 new strong lenses \citep[][hereafter referred to as Papers~I, II, and III, respectively]{huang2020a, huang2021a, storfer2024a} in the DESI Legacy Imaging Surveys \citep{Dey2019} by using residual neural networks trained on observed images.

In this paper, we present results from our fourth search for strong lenses in the Legacy Surveys, now using DR10.
We provide an overview of the observations in \S\ref{sec:observations}.
In \S\ref{sec:model-train}, we describe the construction of the training sample, our neural network ensemble, and our optimized visual inspection method.
New lens candidates are presented in \S\ref{sec:results}.
We discuss our discoveries in \S\ref{sec:discussion} and conclude in \S\ref{sec:conclusion}.

\section{Data}
\label{sec:observations}
\subsection{\textit{The Tractor}}
First, we briefly describe \textit{The Tractor} \citep[][]{lang2016a}, a forward modeling algorithm that performs probabilistic astronomical source detection and typing, and constructs the source catalogs for the Legacy Surveys. 
Source extraction is done on pixel-level data, taking as input the individual images from multiple exposures in multiple bands, with different seeing in each. \textit{The Tractor} treats the fitting process as a $\chi^2$ minimization problem. A detected source is retained if the initial penalized $\chi^2$ is improved by 25\footnote{For more details, see \href{https://www.legacysurvey.org/dr10/description/}{https://www.legacysurvey.org/dr10/description/}}.
\textit{The Tractor} models detected sources as the better of either a point source (``{\tt PSF}'') or round exponential (``{\tt REX}'') galaxy.
A detected source can be further classified as de~Vaucouleurs ({\tt DEV}; S\'ersic index $n=4$) or exponential ({\tt EXP}; $n=1$) profile over {\tt REX}/{\tt PSF} if such a fit improves the $\chi^2$ by 9. 
The classification becomes a S\'ersic profile ({\tt SER}) with an improvement in the $\chi^2$ by another 9 over {\tt DEV}/{\tt EXP}. 

\subsection{Observations}

The Legacy Surveys are composed of three surveys: the Dark Energy Camera Legacy Survey (DECaLS), the Beijing-Arizona Sky Survey (BASS), and the Mayall $z$-band Legacy Survey (MzLS). 
DECaLS is observed by the Dark Energy Camera \citep[DECam;][]{flaugher2015a} on the 4-m Blanco telescope, which covers $\sim 9000$~deg$^2$ of the sky in the range of $-18^{\circ} \lesssim \delta \lesssim +32^{\circ}$. As in DR9, DR10 also contains DECam data reprocessed from the Dark Energy Survey \citep[DES;][]{des2016} for $\delta\lesssim-18^{\circ}$, which provides a footprint of $\sim 5000~$deg$^2$. The newly observed area in DR10 consists of other DECam surveys: the DELVE \citep[][]{{Drlica-Wagner2021}} and DeROSITAS \citep[][]{Zenteno2025} programs. The DECam surveys will hereafter be referred to in their entirety as DECaLS.\footnote{Note that the footprint of DECaLS has expanded from DR7 to DR10, as we show later in this work.}
BASS/MzLS are observed in the $g$ and $r$ bands by the 90Prime camera \citep[][]{williams2004a} on the Bok 2.3-m telescope and in the $z$-band by the Mosaic3 camera \citep[][]{Dey2019} on the 4-meter Mayall telescope. Together, BASS/MzLS cover the same $\sim5000$~deg$^2$ of the northern subregion of the Legacy Surveys. 
Since DR10 includes new imaging data only from DECam, and not in this northern sub-region, \emph{this work performs a lens search only in DECaLS.} 

The Legacy Surveys is imaged with a median $5\sigma$ PSF depths of 23.95 ($g$), 23.54 ($r$), and 22.50 ($z$) AB mag. The median FWHMs for the delivered images are: $1.29\twopr$ ($g$), 1.18$\twopr$ ($r$), and 1.11$\twopr$ ($z$)  for DECaLS \citep[][]{Dey2019}.

Figure~\ref{fig:dr9-footprint} displays the $z$-band depth map of objects classified as {\tt SER} with $z < 20.0$~mag in the Legacy Survey DR10. The figure also outlines the coverage of the DECaLS footprint in DR9, illustrating the overlapping regions as well as the additional areas newly observed in DR10.
{\tt SER} is the most common galaxy type in this magnitude regime. Table~\ref{tab:objects_z20} shows the total counts for each galaxy type ({\tt SER}, {\tt DEV}, {\tt REX}, and {\tt EXP}) with $z$~$<~20.0$~mag for 
the DECaLS region. 

\begin{deluxetable*}{lccccccc}[ht]
\tablewidth{0pt}
\tabletypesize{\scriptsize}
\tablecaption{Object Counts in DECaLS DR10 $z~<~$20 mag \label{tab:objects_z20}}
\tablehead{
\colhead{SER} & & \colhead{DEV} & &  \colhead{REX} & \colhead{EXP} & \colhead{Total by Region}}
\startdata
  $21,482,748$ & & $7,774,926$ & & $9,514,266$ & $4,774,571$ & $43,546,511$\\
 \hline
\enddata
\end{deluxetable*}

\begin{minipage}{\linewidth}
\makebox[\linewidth]{
  \includegraphics[keepaspectratio=true,scale=0.35]{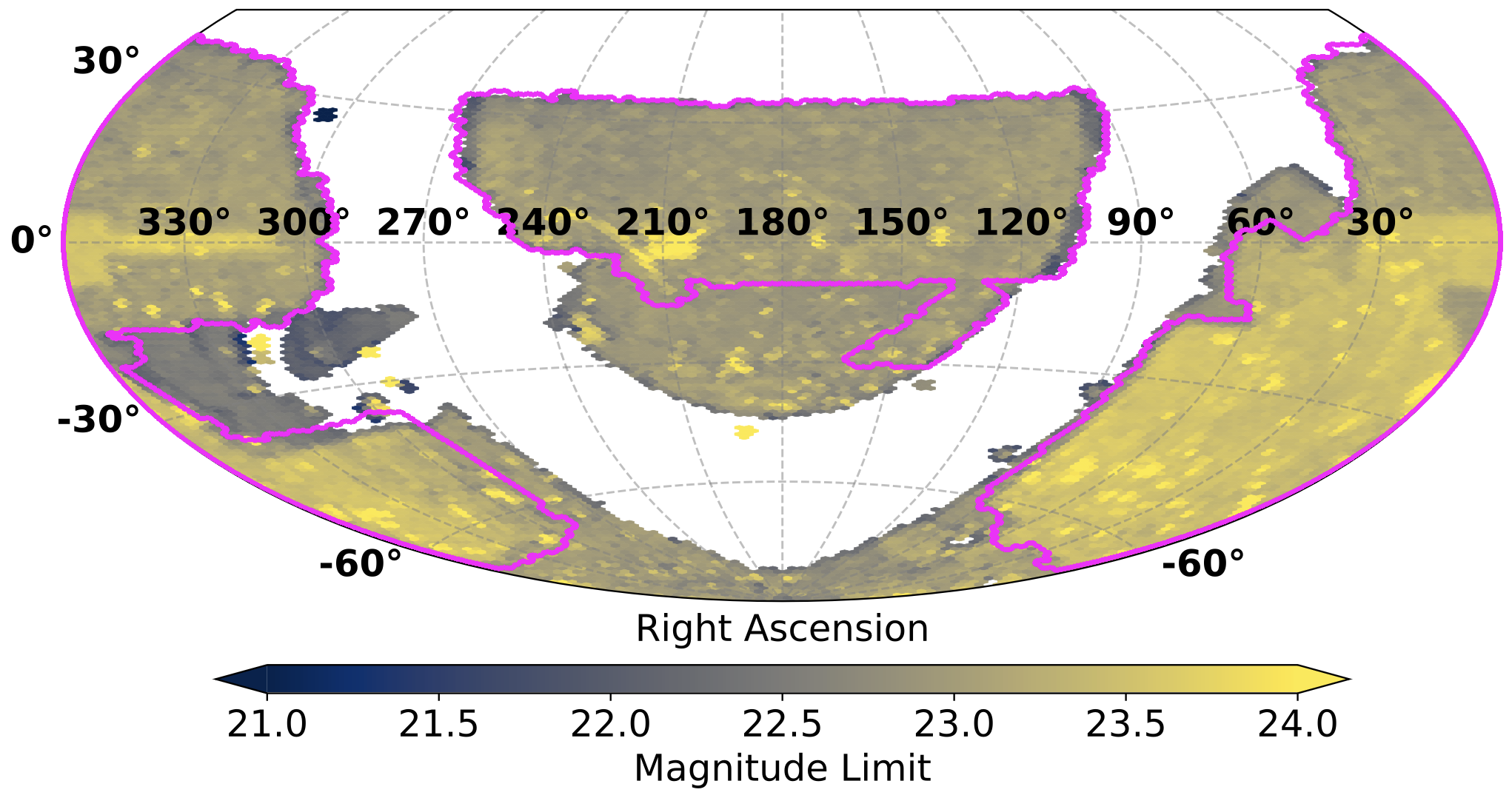}}
\captionof{figure}{\footnotesize The DECaLS survey footprint in DR10 shown in an equal-area Aitoff projection in equatorial coordinates using a hexagonal binning format. The color gradient of the hexagonal bins represents the average $z$-band 5$\sigma$ magnitude limit for galaxies (0.45$^{\prime\prime}$ round exponential profile) within a given bin. These averages are computed from randomly sampled points across the DR10 DECaLS footprint. The DR9 DECaLS footprint is outlined in purple for reference.}

\label{fig:dr9-footprint}
\end{minipage}

\section{The Training Sample, Neural Network Ensemble, and Deployment}
\label{sec:model-train}
\subsection{Training Sample}\label{sec:train}

As in Papers~I, II and III, we continue to use observed images of both lenses and nonlenses for training. Our training sample consists of 1372 lenses and high quality candidates, 869 of which come from Paper~I, II, and III, with the rest selected from the following searches arranged by publication date: \citet[]{zitrin2012,more2012a,anguita2012,brownstein2012a,gavazzi2014a,more2016a,carrasco2017a,diehl2017a,wong2018a,johnson2018,sonnenfeld2018a,sonnenfeld2019a,jacobs2019b,petrillo2019a,jaelani2020a,canameras2020a,sonnenfeld2020a,chan2020a,stein2021a,talbot2021a,rojas2021a,canameras2021a,li2021a,jaelani2021,wong2022a,odonnell2022a,shu2022a}.

To select the lenses for the training sample, we inspect the $\sim 12,000$ known lenses and candidates reported in literature that overlap with the DECam ($\delta<+32^{\circ}$). We apply a uniform set of selection criteria (similar to those in Paper I, II, and III) based solely on the Legacy Surveys images for these systems, regardless of where they were originally discovered. The locations of the selected systems are shown in Figure~\ref{fig:train-lens-map}.

We apply cuts to the training sample based on $z$-band magnitude ($z < 20.0$~mag), location ($\delta$ $<+32^{\circ}$), and depth (at least three exposures\footnote{Papers I-III refer to ``exposures" as ``passes". In this work, we use ``exposures" instead for consistency with Legacy Survey documentation.} in the $g$, $r$, and $z$ bands). The $z$-band magnitude cut was initially tested in Paper I, where we discovered that including objects fainter than the 20th magnitude yields diminishing returns for identifying lens candidates. 

Paper~III excluded {\tt EXP} objects from the training sample since lens galaxies rarely fall under that type. However, the model in Paper~III was able to perform out-of-sample classifications and correctly identified {\tt EXP} candidates during deployment that meet the selection criteria for the training sample in this work. Hence, we now include all types ({\tt SER}, {\tt DEV}, {\tt REX}, and {\tt EXP}) in the training sample. Table~\ref{tab:ts-type-counts} breaks down the counts per \tractor type present in the training sample.

\begin{deluxetable*}{lccccccc}[h]
\tablewidth{0pt}
\tabletypesize{\scriptsize}
\tablecaption{Lenses Counts 
in the Training Sample
\label{tab:ts-type-counts}}
\tablehead{
\colhead{SER} & & \colhead{DEV} & &  \colhead{REX} & \colhead{EXP}}
\startdata
 \hline 
 960& &266& &112 &34 \\ 
\enddata
\end{deluxetable*}
\begin{minipage}{\linewidth}
\makebox[\linewidth]{
  \includegraphics[keepaspectratio=true,scale=0.6]{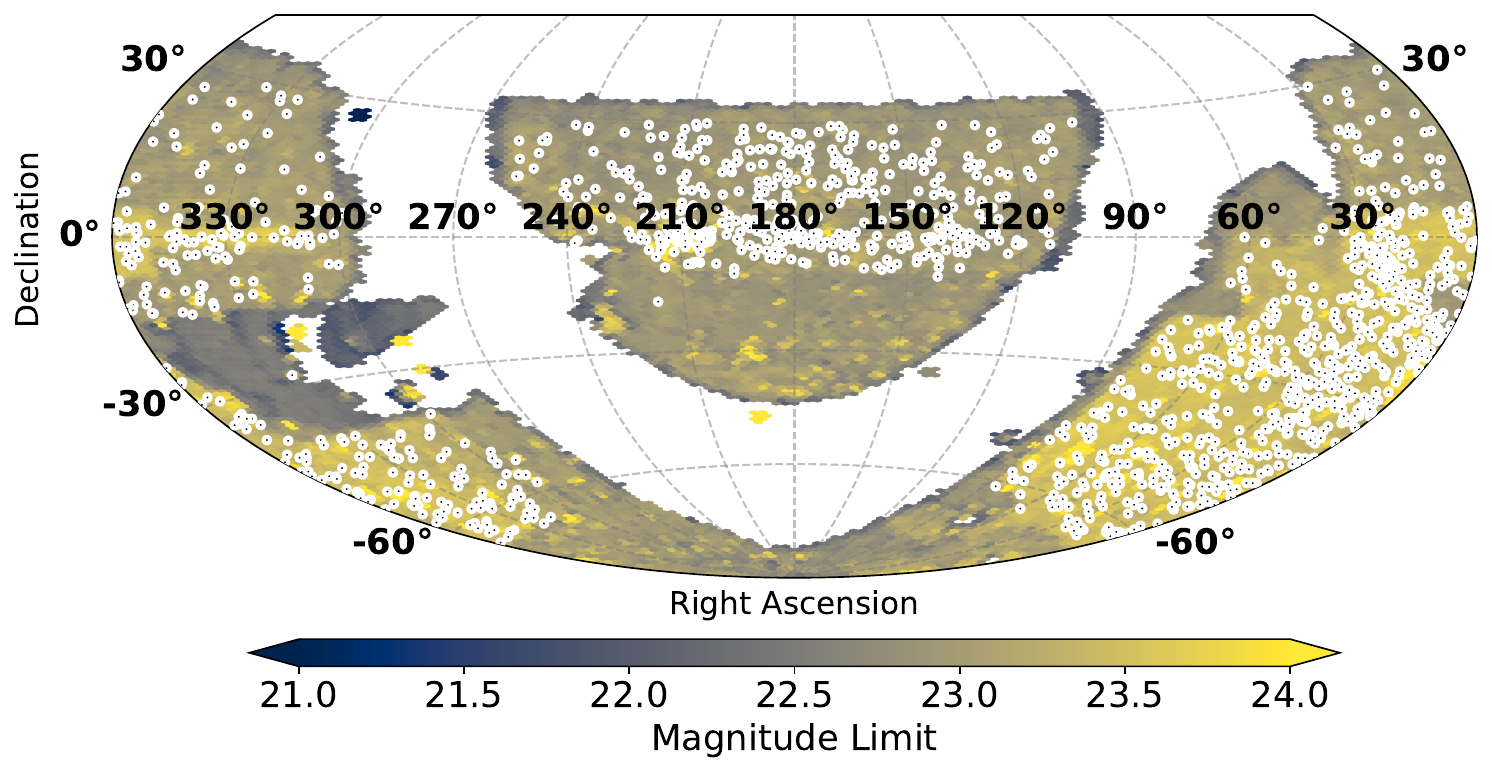}}
\captionof{figure}{\footnotesize The 1372 lenses included in the training sample over the depth map of the Legacy Surveys DR10 shown in Figure~\ref{fig:dr9-footprint}.}
\label{fig:train-lens-map}
\end{minipage}

For the selection of nonlenses, for each type ({\tt SER}, {\tt DEV}, {\tt REX}) in DECaLS, we bin by the number of exposures in $z$-band. This is to prevent potential bias by the neural net based on the number of exposures (see Paper~II). We then select nonlenses randomly, but keep the proportionality to the lenses the same in each bin ($\sim$~100:1). This ratio has been steadily increased from Paper~I through Paper~III (where this ratio was $\sim$~33:1 in Paper III) to better reflect the expected nonlens-to-lens ratio during deployment, which is estimated to be $\sim$~10,000:1.
We cross-reference this nonlens selection with the $\sim12,000$ known lenses and candidates found in the literature to remove potential contaminants (i.e., cutouts that contain a lens or a lens candidate). This is made possible implementing \texttt{spherimatch}\footnote{\href{https://github.com/technic960183/spherimatch}{https://github.com/technic960183/spherimatch}} \citep[][]{Hsu2025}, a package intended for cross-referencing catalogs in spherical coordinates. The package has a time complexity of  $\mathcal{O}(N \log N)$, making it well-suited for handling large catalogs, as required in this work. In an effort to remove possible additional contamination, we deploy the trained ResNet from Paper III to this selection and inspect the images that receive a probability greater than 0.4 (the same threshold used in Paper III). After visually inspecting the recommendations from this model, we remove 9 potential contaminants from this selection. The contaminants we remove are comparable to or better than Grade D in Paper~III (also, see \S~\ref{sec:VI-scheme}). This results in a total of 134,182 nonlenses in the training sample. 
This is 10$\times$ larger than training sample in our first lens search (Paper~I).

The final dataset of lenses and nonlenses is randomized and divided into 70\% for training and 30\% for validation, while maintaining the $\sim$~100:1 nonlens-to-lens ratio in each split. For both the training and validation catalogs, we extract image cutouts of 101 × 101 pixels, corresponding to an angular size of $\sim$~26$^{\prime\prime}$ × 26$^{\prime\prime}$. The exact same training and validation sets are used to train our two models described below in \S~\ref{sec:model}.

\subsection{The Neural Network Models}\label{sec:model}

In this section, we describe the two neural network models used in our lens search, a residual network (ResNet; \S~\ref{sec:shielded}) and a so-called ``EfficientNet'' (\S~\ref{sec:efficientnet}).

\subsubsection{ResNet}\label{sec:shielded}
The ``shielded'' residual network (ResNet) model, as described in Paper~II, is an improvement upon the architecture presented in \citet[][L18]{lanusse2018a}.
The ``shielding'' layers \citep{szegedy2014a} perform 1$\times$1 convolutions that reduce the dimensionality (i.e., the number of layers) of the output from each ResNet block.

We use the cross-entropy loss function,

\begin{linenomath*}
\begin{equation}
\label{eqn:loss}
    \displaystyle\mathcal{L}_{CE} = -\sum_{i=1}^{N} y_i \log \hat{y}_i+(1-y_i) \log (1-\hat{y}_i)
\end{equation}
\end{linenomath*}

\noindent
where $y_i$ is the label for the $i$th image (1 for lens and 0 for nonlens), $\hat{y}_i \in [0,1]$ is the model predicted probability, and $N$ is the number of images in one training step (the same as batch size, given below). 
During training, the ResNet attempts to minimize the loss while the Area Under the Curve (AUC) metric is used throughout training to identify the best-performing model. Figure~\ref{fig:loss_roc_S} shows the loss minimization process alongside the Receiver Operating Characteristic (ROC) curve for the model that achieved the highest validation AUC during training.
 
The ResNet model contains a total of 194,433 trainable parameters and was trained with a batch size of 2048 ($\times$4 more than the EfficientNet, \S~\ref{sec:efficientnet}), an initial learning rate (LR) of ~5$\times 10^{-4}$, and a learning rate decay schedule that reduced the LR by half at the 80th epoch. 
The ResNet is trained on 4 NVIDIA A100 GPUs on one GPU node on the \emph{Perlmutter} supercomputer at the National Energy Research Scientific Computing Center (NERSC)\footnote{A U.S. Department of Energy Office of Science user facility, \url{https://www.nersc.gov/}. }.
 Using TensorFlow's MirroredStrategy, training on 4 GPUs is straightforward. The multi-GPU implementation results in a speed gain of $3.7\times$ (remarkably close to the naive expectation of $4\times$): the training time per epoch is reduced from 78 to 21 seconds.

In addition to the cross-entropy loss, we further assess the performance of our trained ResNet using the ROC, which plots the True Positive Rate (TPR) against the False Positive Rate (FPR), where a “Positive” (P) indicates a lens and a “Negative” (N) a nonlens. The definitions of TPR and FPR are given as:

\begin{linenomath*}
\begin{equation}
\label{eqn:TPR}
\rm{TPR} = \frac{\rm{TP}}{\rm{P}} = \frac{\rm{TP}}{\rm{TP} + \rm{FN}}
\end{equation}
\end{linenomath*}

and

\begin{linenomath*}
\begin{equation}
\label{eqn:FPR}
\rm{FPR} = \frac{\rm{FP}}{\rm{N}} = \frac{\rm{FP}}{\rm{FP} + \rm{TN}}
\end{equation}
\end{linenomath*}
where True Positives (TP) are lenses correctly identified, False Positives (FP) are nonlenses incorrectly identified as lenses, True Negatives (TN) are nonlenses correctly rejected, and False Negatives (FN) are lenses incorrectly rejected.

Random classifications produce a diagonal line on the ROC curve with an AUC of 0.5, while a perfect classifier achieves an AUC of 1. During training, the validation AUC is calculated at each epoch, and the model with the highest validation AUC is selected for deployment. 

\begin{minipage}{\linewidth}
\makebox[\linewidth]{
  \includegraphics[keepaspectratio=true,scale=0.7]{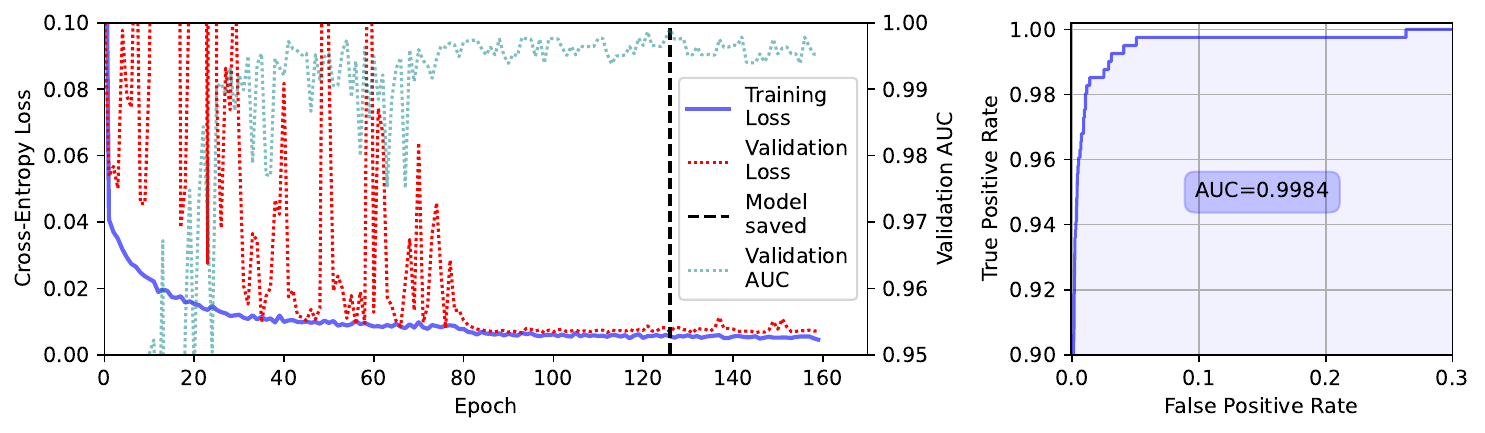}}
\captionof{figure}{\footnotesize Left: Cross-entropy loss for the training and validation sets (left y-axis) and validation AUC (right y-axis) vs.\ training epochs. The vertical dashed black line at epoch 126 marks the point where the model achieved its highest AUC at 0.9984. This trained model was selected for deployment.
Right: ROC curve for the validation set using the best-performing model from epoch 126.}
\label{fig:loss_roc_S}
\end{minipage}

\subsubsection{EfficientNet Model}\label{sec:efficientnet}
For the first time with respect to Papers I through III, we integrate the EfficientNetV2, referred to in this work as the ``EfficientNet,'' which improved the EfficientNetV1 presented in \cite{Tan2019}. EfficientNetV1 comes from the search for a principled method for scaling Convolutional Neural Network (CNN) parameters (width of layers, depth of layers, resolution scaling) in order to achieve high accuracy, even with a small model that can be run on mobile devices. The primary building block for EfficientNetV1 is the mobile inverted bottleneck MBConv from \cite{Sandler2018}, with a squeeze and excitation layer. The MBConv block's input and output are thin ``bottle-necked'' layers as opposed to the standard residual block, which uses expanded representations in the input. In an MBConv block, the input is first expanded with a 1$\times$1 convolution which increases the number of channels. Lightweight depthwise convolution is performed on each channel of the expanded input separately, followed by a pointwise (1$\times$1) convolution, which builds features via linear combinations of channels. The output is sent into a squeeze-and-excitation block, an attention mechanism which gives weights to each feature map. The squeeze and excitation block's output is put through another pointwise (1$\times$1) convolution, this time to reduce the number of channels back down (This is similar to the ``shielding'' layers we added to the ResNet model; \S~\ref{sec:shielded}). Compared to a standard convolution, this process requires less memory and fewer computations. However, because depthwise convolutions suffer from poor parallelization, they are slower to train using modern GPUs. To improve speed, EfficientNetV2 introduces Fused-MBConv blocks, which replace the expansion and depthwise convolution with a regular 3$\times$3 convolution. While this is more expensive in terms of parameters and computations, because of parallelization, using FusedMBConv in early layers results in a $\sim$40\%  increase in training speed when using a GPU, and only a $\sim$4\% increase in parameters compared to EfficientNetV1.

Unlike our ResNet model, the EfficientNet was pre-trained. 
In this work, we fine-tune the EfficientNet model using the same cross-entropy loss function as for the ResNet (Equation~\ref{eqn:loss}).
The training results are shown in Figure~\ref{fig:loss_roc_E}.

The EfficientNet model contains a total of 20,542,883 trainable parameters and was trained with a batch size of 512, an initial learning rate (LR) of 3.88 $\times 10^{-4}$, and a learning rate decay schedule that reduced the LR by half at the 130th epoch. Training was also conducted on 4 GPUs at NERSC (the same as in \S~\ref{sec:shielded}).
With a per GPU batch size of 512, the training time per epoch improves from 72 to 23 seconds, or, an approximately $3\times$ speed gain, slightly less than the naive expectation of 4$\times$. This results in a 3 hour reduction of training time, from 4.7 hours to train a 160 epoch model down to 1.3 hours.

Figure~\ref{fig:loss_roc_E} shows the AUC and loss progression for the training and validation sets, and the ROC curve for our best-performing model, with an AUC of 0.9987, which is comparable to the ResNet's best AUC. 

\begin{minipage}{\linewidth}
\makebox[\linewidth]{
  \includegraphics[keepaspectratio=true,scale=0.7]{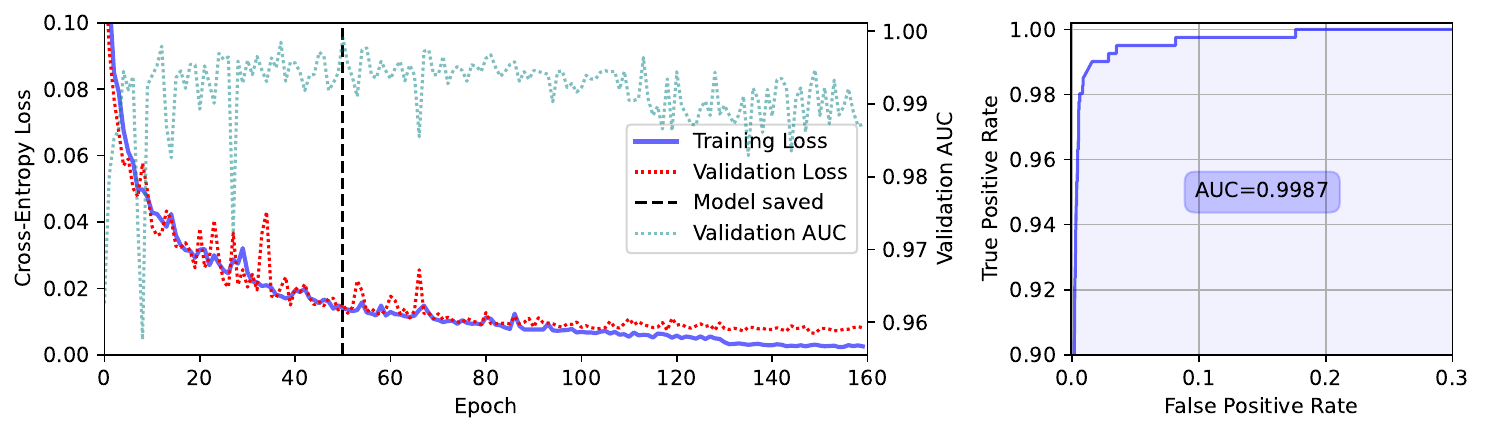}}
\captionof{figure}{\footnotesize Left: Cross-entropy loss for the training and validation sets (left y-axis) and validation AUC (right y-axis) vs.\ the training epochs. The vertical dashed black line at epoch 50 marks the point where the model achieved its highest AUC. This trained model was selected for deployment.
Right: ROC curve for the validation set using the best-performing model from epoch 50. }
\label{fig:loss_roc_E}
\end{minipage}

The divergent behavior of the loss and AUC curves in Figure~\ref{fig:loss_roc_E} is worth noting. After epoch 50, where the model achieves the highest validation AUC, the training and validation loss continue to improve (i.e., decrease), unlike the validation AUC, which deteriorates. The validation AUC fluctuates near 0.995 before starting to decline around epoch 80. Figure~\ref{fig:eff-hist-comparison} addresses this discrepancy by comparing the probability distributions of an EfficientNet at 50 epochs vs. at 160 epochs, showing that the 160-epoch model puts 283 more images in its smallest prediction probability bin (0, 0.04) than the 50-epoch model. 
Due to the significant class imbalance in the training sample (negative examples outnumber positive examples by 100:1), the model is able to achieve an overall lower loss by pushing these images into the lowest probability bin. As indicated by the lower panel of Figure~\ref{fig:auc-comparison}, most of the additional 283 images came from images assigned a probability in the adjacent (low) probability bins. 
The vast majority of images that were not in the top or bottom bin (predicted probability between 0.04 and 0.96) were already correctly identified as TNs by the 50-epoch model (300 of 404 images were TNs). Thus, pushing them into the lowest bin improves (lowers) the loss but does not improve the AUC.
On the other hand, 104 of these images are TPs, so pushing them to the lowest bin \emph{should} negatively affect our performance. The lowest bin of the 50-epoch model is 99.98\% TNs, but the same bin of the 160-epoch model is only 99.94\% TNs. This reduction in purity will increase the loss. However, given the much larger number of TNs that got pushed into the lowest bin, the overall effect of redrawing the boundary by the 160-epoch model will improve (lower) the loss, but hurt (lower) the AUC.
Similarly, some lenses that were assigned to the second highest probability bin by the 50-epoch model are now assigned to the highest probability bin by the 160-epoch model. This will improve (lower) the loss but will not affect the AUC.
This logic is why the AUC is the definitive metric to select the best-performing model for deployment.

\vspace{0.5cm}
\begin{minipage}{\linewidth}
\makebox[\linewidth]{
  \includegraphics[keepaspectratio=true,scale=0.7]{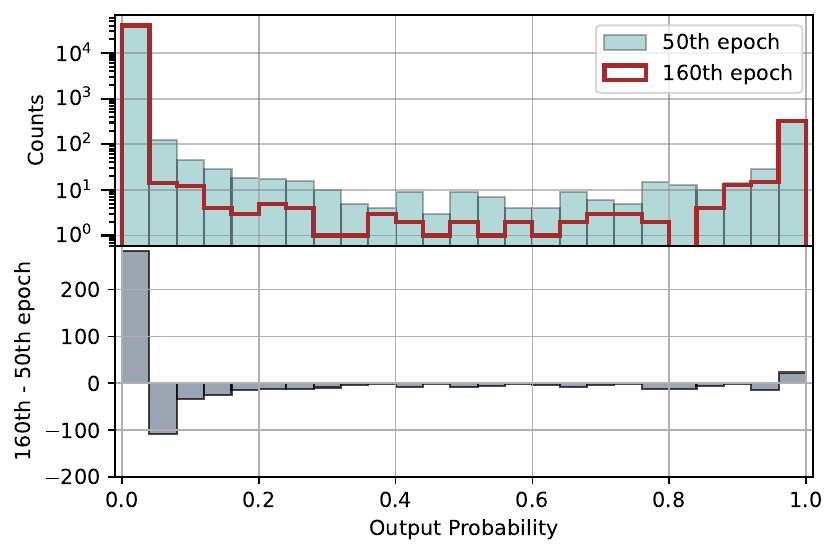}}
\captionof{figure}{\footnotesize Upper panel: The probability distributions on the validation set from the EfficientNet at the 50th epoch (clear blue) and at the 160th epoch (maroon outline), where the counts are shown in logarithmic scale. Lower panel: differences between counts at the 160th and 50th epoch (160th epoch - 50th epoch counts) in linear scale. The histogram shows a noticeable drop in the bins between the first and last in the 160-epoch model compared to the 50-epoch model (note that these differences are difficult to see in the upper panel, given its logarithmic scale). Many images that received middle-range predictions from the 50-epoch model are pushed into the lowest bin by the 160-epoch model as it tries to minimize the loss. But this actually decreases the AUC and therefore makes the performance worse (see text).
To a much lesser extent, a similar trend can be observed for the highest probability bin, which will improve (lower) the loss but not affect the AUC.}

\label{fig:eff-hist-comparison}
\end{minipage}

\subsubsection{The Meta-learner}\label{sec:meta-learner}
The ResNet and EfficientNet models, described in \S~\,\ref{sec:shielded} and \S~\ref{sec:efficientnet}  respectively, generate independent predictions for each image during deployment. To combine these predictions, we employ feature weighted stacking \citep[]{COSCRATO2020141}, which consists of a meta-learner that aggregates the probabilities produced by the two base models. This meta-learner is implemented as a simple one-layer neural network with 300 nodes, designed to take the probabilities of the base models as its input and output a single probability.

The meta-learner is trained using the probability outputs of the base models on the training dataset. During validation, the meta-learner is evaluated by feeding it the predictions of the base models on the validation set. The primary objective of the meta-learner is to 
find an optimal
combination of the probabilities of the two base models.

\vspace{0.5cm}
\begin{minipage}{\linewidth}
\makebox[\linewidth]{
  \includegraphics[keepaspectratio=true,scale=0.65]{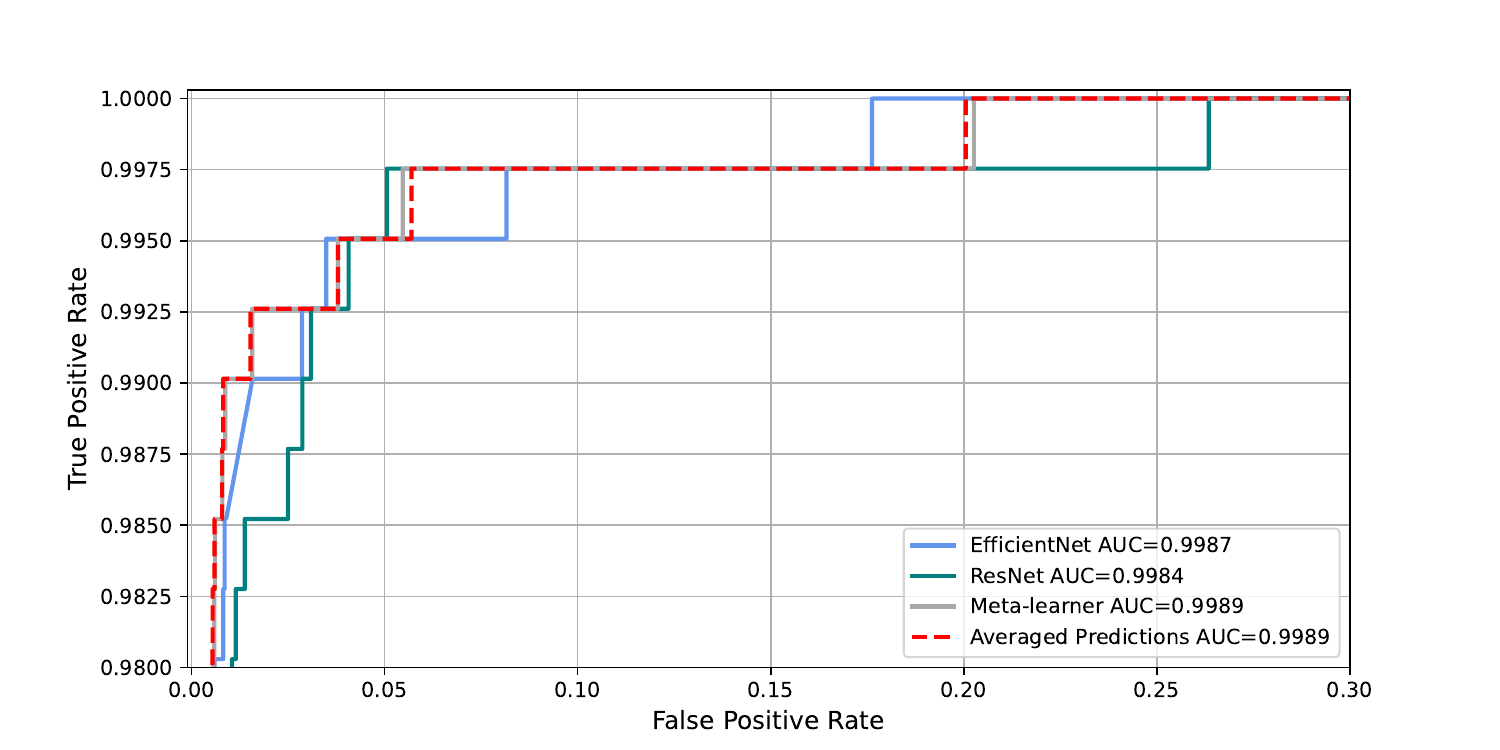}}
\captionof{figure}{\footnotesize ROCs for the base models individually (ResNet and EfficientNet), the meta-learner, and simple averaging of the base models' predictions. The legend displays the AUC for each of these ROCs.}
\label{fig:auc-comparison}
\end{minipage}

Given that the two base models are trained on the same dataset, their predictions are inherently correlated. This correlation constrains the meta-learner’s ability to substantially outperform simple averaging of the base models’ predictions, as illustrated by the AUCs in Figure~\ref{fig:auc-comparison}. However, the meta-learner still optimizes the weighting of the base model contributions, which provides a more systematic approach than manual averaging.
Figure~\ref{fig:auc-comparison} shows that the meta-learner achieves a higher overall AUC compared to each individual base model.

To further enhance the use of two or more models, strategies such as training the base models on different subsets of the data could be explored, as doing so may produce more diverse predictions by the base models. 
For example, \cite{shu2022a} trained models using lenses at different redshifts, which resulted in the models finding different candidates. 

\subsection{Visual Inspection}\label{sec:VI}
\subsubsection{Visual Inspection Scheme}\label{sec:VI-scheme}

We choose a probability threshold of $\geq 0.9867$ from the meta-learner, resulting in \Ninspectedsystems recommendations from approximately 43 million images (the top 0.01 percentile). Out of \Ninspectedsystems pre-selected images, we remove 1102 recommendations present in the $\sim$12,000 known candidates and systems from literature.
Our chosen probability threshold and elimination of known systems from visual inspection ensures that only the most confident predictions are considered and only \textit{new} discoveries are reported in this paper.
This final set of 4578 recommendations is then visually inspected to identify small blue or red galaxy/galaxies next to  (typically) a red galaxy/galaxies that exhibit the following traits:
    \begin{itemize}
        \item are typically 1 - 5$\twopr$ away from the red galaxy/galaxies
        \item have low surface brightness
        \item curve toward the red galaxy/galaxies
        \item have counter/multiple images with similar colors (especially in Einstein-cross like configuration)
        \item are elongated (including semi- or nearly full rings)
    \end{itemize}
\noindent
We assign grades based on the number of traits and the level of clarity of each trait exhibited by a recommendation. 
The criteria used by human graders in this work, built upon  the ones used in Papers~I through III, are the following: 
\begin{itemize}
    \item Grade~A:  We have a high level of confidence in these candidates. Many of them have one or more prominent arcs, usually blue.
    The rest have one or more clear arclets, sometimes arranged in multiple-image configurations with similar colors, also typically blue. However, there are clear cases with red arcs as well.
    
    \item Grade~B: Giant arcs tend to be fainter than those for Grade A. Likewise, the putative arclets tend to be smaller and/or fainter, or isolated (without counter images).
    
    \item Grade~C: They generally have features that are even fainter and/or smaller than what is typical for Grade~B candidates, but are nevertheless suggestive of lensed arclets. Counter images are often not present or hardly discernible. In a number of cases, the angular scales of the candidate systems are comparable to or only slightly larger than the seeing. Therefore, to attain a higher level of certainty for some of these candidates, higher spatial resolution, deeper data, or spectroscopic observations would be required.
\end{itemize}
Co-authors JI, BK, CS, and XH performed generous initial inspection of the recommended images using the grading scheme above. CS and XH provided final inspection grades, and we report the average and difference of the results of these two graders.

\subsubsection{Visual Inspection Portal}\label{sec:VI-portal}
To optimize the grading process, 
we have developed a visual inspection portal. Given a list of recommendations, the portal displays each co-added image along with its individual \textit{grz} channels with ``Previous"/``Next" buttons that enable scrolling through individual recommendations. 
The cutout image for each recommendation is displayed with text-boxes to enter a grade and an optional comment, which are written into a CSV file. 
Each image is also linked to the Legacy Surveys' sky viewer\footnote{\href{https://www.legacysurvey.org/viewer}{https://www.legacysurvey.org/viewer}} for zoom and other options, e.g., to provide larger scale context for a possible group/cluster scale lens. Figure~\ref{fig:VI_demo} shows the user interface of the portal.

\begin{minipage}{\linewidth}
\makebox[\linewidth]{
  \includegraphics[keepaspectratio=true,scale=0.55]{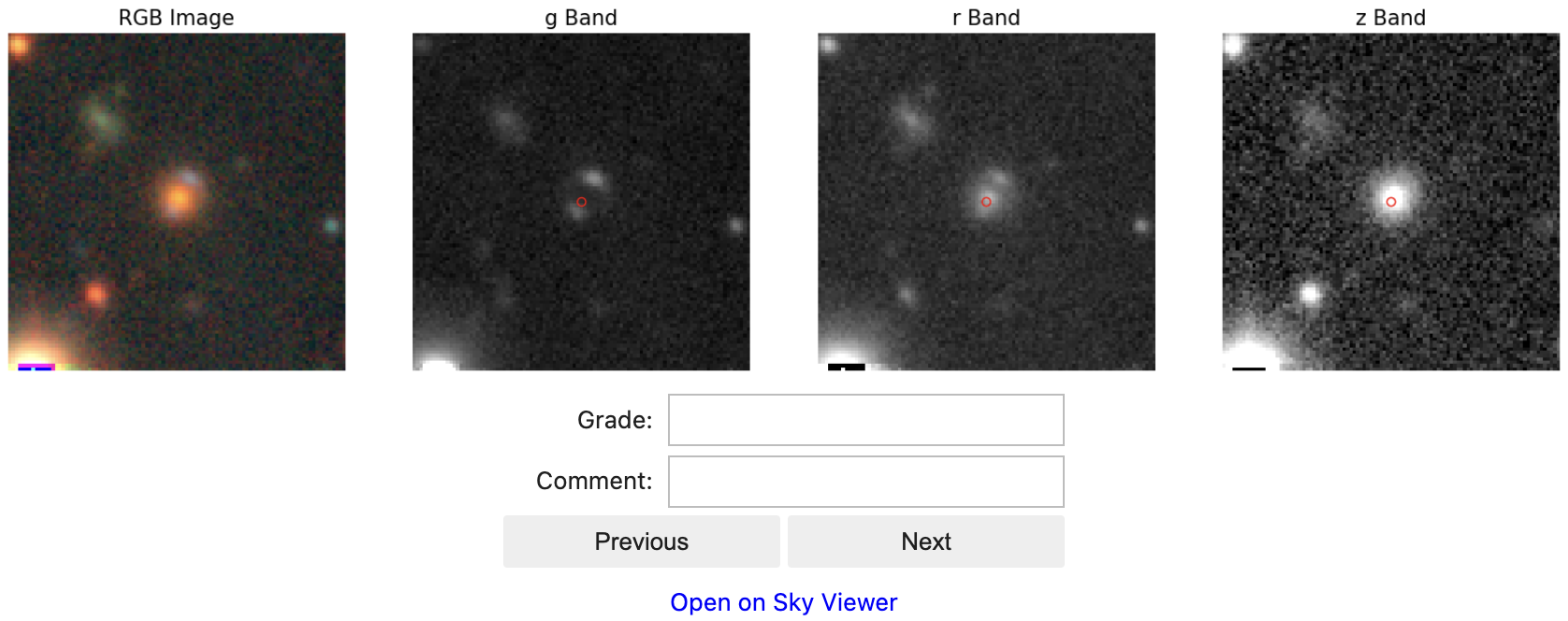}}
\captionof{figure}{\footnotesize User interface of the visual inspection portal, displaying images from left to right: co-added image, and the \textit{g}, \textit{r}, and \textit{z} bands. The candidate system DESI-316.3086-07.3386, identified for the first time on this paper, is shown as an example. In this case, the arc, and especially the counter-arc are more clearly visible in the \textit{g}-band without the putative lens (the red galaxy at the center of the cutout, highlighted by the red circle).}

\label{fig:VI_demo}
\end{minipage}

With respect to Papers I–III, this is the first time that individual \textit{grz} bands are used in addition to the co-added images to visually inspect recommendations. 
Typical lenses are early-type galaxies with redder spectral energy distributions (SEDs), often peaking at longer wavelengths, which makes them more prominent in the \textit{z} band. Conversely, lensed sources, which are often star-forming galaxies or quasars, tend to have bluer SEDs, making their arcs or multiple images more prominent in the \textit{g} band. The \textit{r} band provides an intermediate perspective, capturing both the redder lenses and bluer sources with moderate contrast. By examining the separate bands, it is possible to detect subtle features that may otherwise be overlooked in a combined RGB image, as it has been noted by other lens searches such as \cite{storfer2025}.

\section{Results}
\label{sec:results}
In this section, we first present the lens candidates identified from the visual inspection in this work (\S\,\ref{sec:dr9-cands}).
We then provide a summary of all lens candidates discovered by our group in the Legacy Surveys DR7, 8, 9, and 10  (\S\,\ref{sec:all-cands}). 
\subsection{Lens Candidates in DR10}\label{sec:dr9-cands}
In this section, we present the strong lens candidates discovered in Legacy Surveys DR10 from this work. We deployed our trained models on $\sim$~43 million cutouts with $\geq 3$ exposures in $g$, $r$ $z$-bands, centered on all galaxy types (non-{\tt PSF}) with $z$-band mag $< 20.0$~mag in the DECaLS footprint.

As described in Section~\ref{sec:VI-scheme}, the top \Ninspectedsystems recommendations with highest probabilities are selected, and then 1102 known candidates and systems are removed, which come from:
\citet[]{stein2021a,odonnell2022a,jaelani2020a,canameras2020a,petrillo2019a,huang2020a,huang2021a,wong2022a,sonnenfeld2018a,wong2018a,li2021a,sonnenfeld2019a,rojas2021a,jacobs2019a,talbot2021a,storfer2024a,more2016a,more2012a,sonnenfeld2020a,diehl2017a,gavazzi2014a,savary2022a,wong2022a,canameras2021a,carrasco2017a,chan2020a}

Resulting from visual inspection, we report \NnewlensesUndisc \textit{new} lens candidates. These candidates contain \NnewlensesUndiscA Grade As, \NnewlensesUndiscB Bs, and \NnewlensesUndiscC Cs. 
Figure~\ref{fig:example-cands} shows three examples for each of the four categories of new lens candidates. The locations in the sky of all new candidates are shown in Figure~\ref{fig:cand_map_dr9} and Table~\ref{tab:shielded-cands} breaks them down by \textit{The Tractor} type.

\begin{deluxetable*}{lccccccc}[h]
\tablewidth{0pt}
\tabletypesize{\scriptsize}
\tablecaption{Strong Lens Candidates by Type\label{tab:shielded-cands}}
\tablehead{\colhead{Grade} & \colhead{A} & & \colhead{B} & &  \colhead{C} & \colhead{Total by Type}}
\startdata
    SER & 65& &71& &357 & 493 \\
    DEV & 9& &12& &100 & 121 \\
    REX & 11& &18& &133 & 162 \\
    EXP & 5& &3& &27 & 35 \\
 \hline
 Total by Grade & 90& &104& &617 & 811 \\
\enddata
\end{deluxetable*}
\vspace{-1.3cm}
\begin{minipage}{\linewidth}
\makebox[\linewidth]{
  \includegraphics[keepaspectratio=true,scale=0.5]{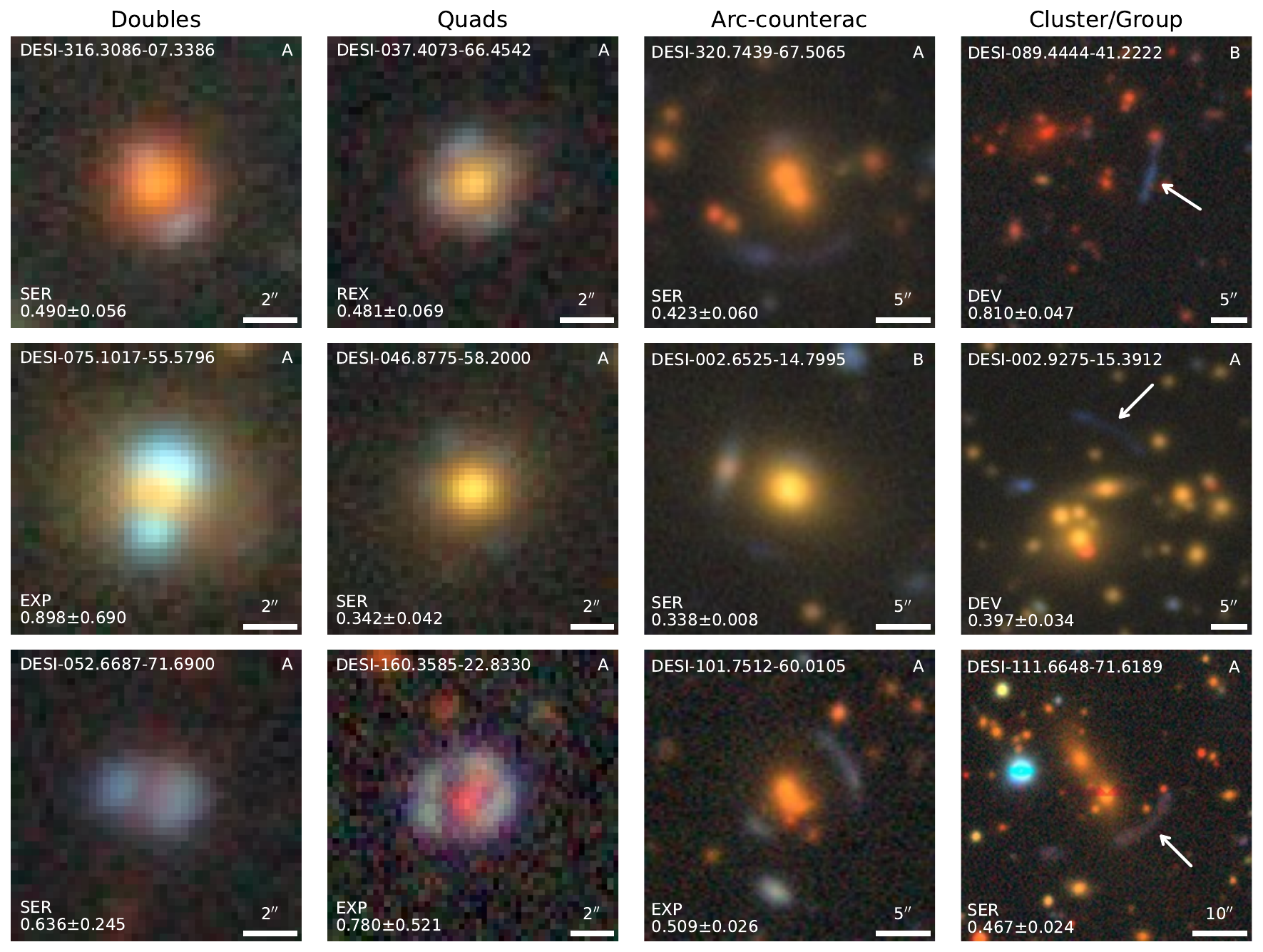}}
\captionof{figure}{\footnotesize Twelve of the \NnewlensesUndisc new candidates found in this work, categorized into four groups: doubles, quads, arc-counterarc systems, and cluster/group lenses. The naming convention follows R.A. and decl. in decimal format. All images have north up and east to the left. 
The grade is shown at the top-right corner of each image, while the bottom-left corner indicates 
\textit{The Tractor} type and the photometric redshift of the lens ($z_d$). \textbf{First column:} doubly lensed systems. \textbf{Second column:} quadruply lensed systems. \textbf{Third column:} systems with an arc–counterarc pair. \textbf{Fourth column:} groups/cluster lenses, where putative arcs are indicated with white arrows.}
\label{fig:example-cands}
\end{minipage}

Our search demonstrates that we can find a large number of 
new strong lens candidates in a footprint that has been mined for lenses before in certain sub-regions, repeatedly. Remarkably, $\sim60\%$ of our 811 new candidates lie within the DR9 footprint, which is an area that has been the focus of multiple previous lens searches. Table~\ref{tab:dr9-dr10-comparison} breaks down the distribution of these candidates between the previously searched DR9 region and the newly added area in DR10. Figure~\ref{fig:candidates-dr9} shows a subset of the highest-quality (A-grade) candidates found within DR9. The abundance of new high-quality candidates from DR9 in this work may be owing to our use of separate bands for visual inspection (see \S\ref{sec:VI-portal}), which previous searches within DR9 lack.

\begin{deluxetable*}{lccccccc}[h]
\tablewidth{0pt}
\tabletypesize{\scriptsize}
\tablecaption{Strong Lens Candidates\label{tab:dr9-dr10-comparison}}
\tablehead{\colhead{Grade} & \colhead{A} & & \colhead{B} & &  \colhead{C} & \colhead{Total by Type}}
\startdata
    Within DR9 & 21& &51& &412 & 484 \\
    Outside DR9 & 69& &53& &205 & 327 \\
 \hline
 Total by Grade & 90& &104& &617 & 811 \\
\enddata
\end{deluxetable*}

\begin{minipage}{\linewidth}
\makebox[\linewidth]{
  \includegraphics[keepaspectratio=true,scale=0.45]{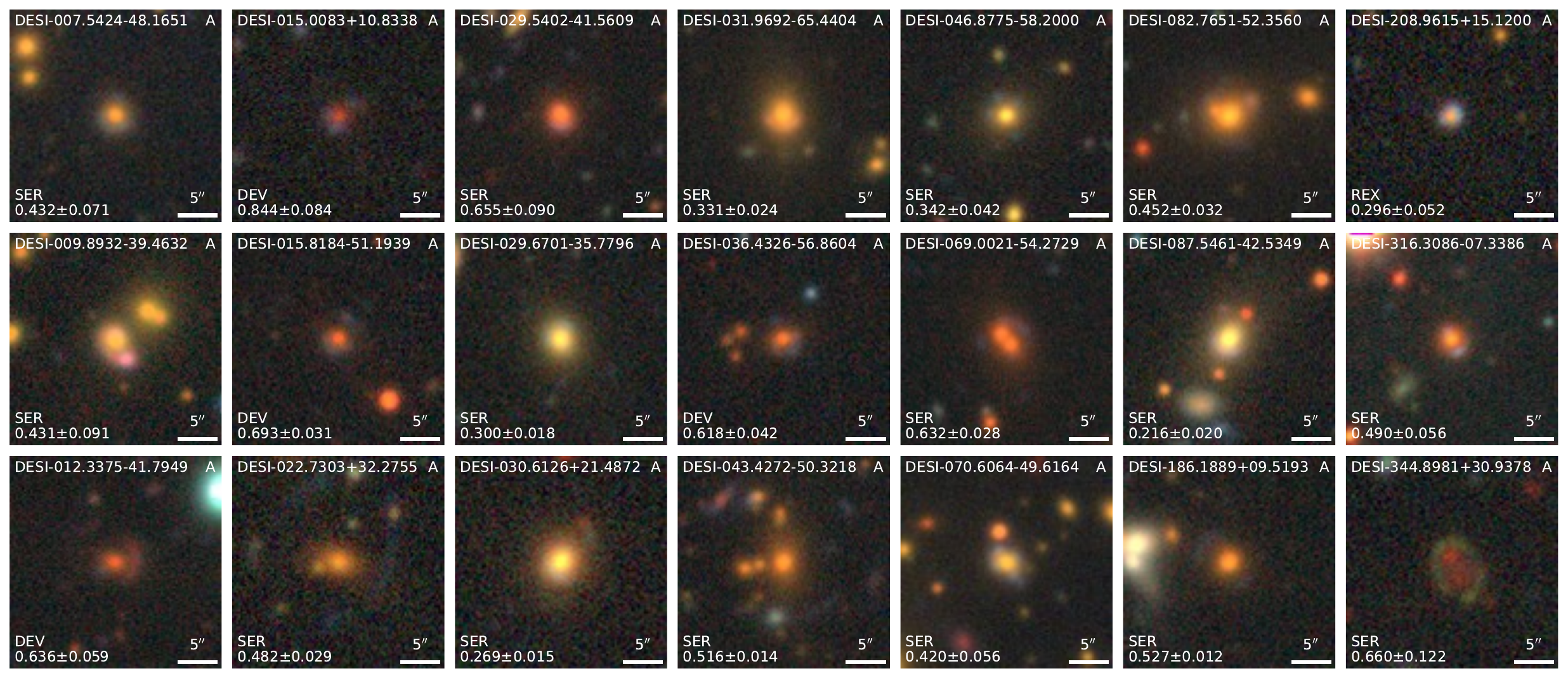}}
\captionof{figure}{All 21 newly identified A-grade strong lens candidates located within the DR10 data but falling inside the previously mined DR9 footprint.}
\label{fig:candidates-dr9}
\end{minipage}

The purity of systems by type is shown in Table~\ref{tab:nn-purity}. These values are computed by scaling the grade distribution of \textit{new} candidates to known ones, under the assumption that both new and rediscovered candidates follow the same grade distribution. This is due to the fact that only recommendations that did not match with known candidates and systems were inspected. The {\tt SER}, {\tt DEV}, and {\tt REX} types exhibit improved purity compared to Paper III, where, respectively, one in every 23, 22, and 27 recommendations was a candidate. The only exception to this trend is galaxies typed as {\tt EXP} by \textit{The Tractor}, for which Paper III reported a purity of 39, whereas this work finds a purity of 80.

\begin{deluxetable*}{lcccccccc}[h]
\tablewidth{0pt}
\tabletypesize{\scriptsize}
\tablecaption{Purity for Our Neural Network Ensemble\label{tab:nn-purity}}
\tablehead{\colhead{\tractor Type} & \colhead{SER} & & \colhead{DEV} & &  \colhead{REX} & \colhead{EXP}}
\startdata
 \hline
 Purity by Type&  \SERpurity& &\DEVpurity& &\REXpurity &\EXPpurity \\ 
\enddata
\tablecomments{Purities are shown as the number of Neural Network Ensemble recommendations inspected in order to find a lens candidate.}
\end{deluxetable*}

\begin{minipage}{\linewidth}
\makebox[\linewidth]{
 \includegraphics[keepaspectratio=true,scale=0.6]{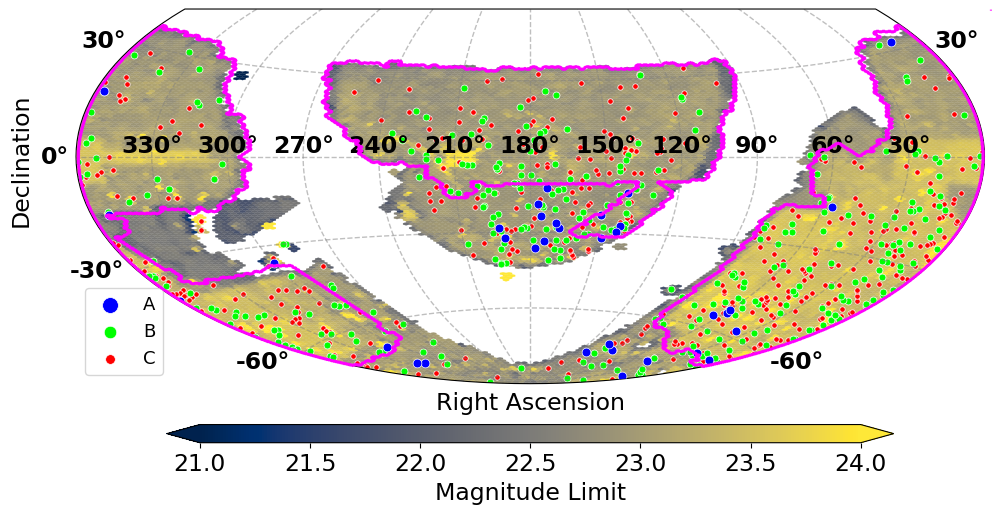}}
\captionof{figure}{\footnotesize The \NnewlensesUndisc new candidate lensing systems discovered in this work by grades 
over the depth map of the Legacy Surveys DR10 shown in Figure~\ref{fig:dr9-footprint}. The magenta contours enclose the DECaLS region of DR9 for comparison.}
\label{fig:cand_map_dr9}
\end{minipage}

\subsection{Lens Candidates in DR7, 8, 9, and 10}\label{sec:all-cands}

From our four searches in the Legacy Surveys, we have found a grand total of \grandtotnew \textit{new} candidates (Paper~I in DR7 among {\tt DEV} and {\tt COMP} in DECaLS only, Paper~II in DR8, Paper~III in DR9, and this work in DR10). The entire catalog of these lenses can be found on our project website\footnote{\href{https://sites.google.com/usfca.edu/neuralens/publications/lens-candidates-inchausti-et-al-2025}{https://sites.google.com/usfca.edu/neuralens/publications/lens-candidates-inchausti-et-al-2025}}. The positions of all candidates on the sky are shown in Figure~\ref{fig:cand_map_all}.

\vspace{0.1cm}
\begin{minipage}{\linewidth}
\makebox[\linewidth]{
  \includegraphics[keepaspectratio=true,scale=0.6]{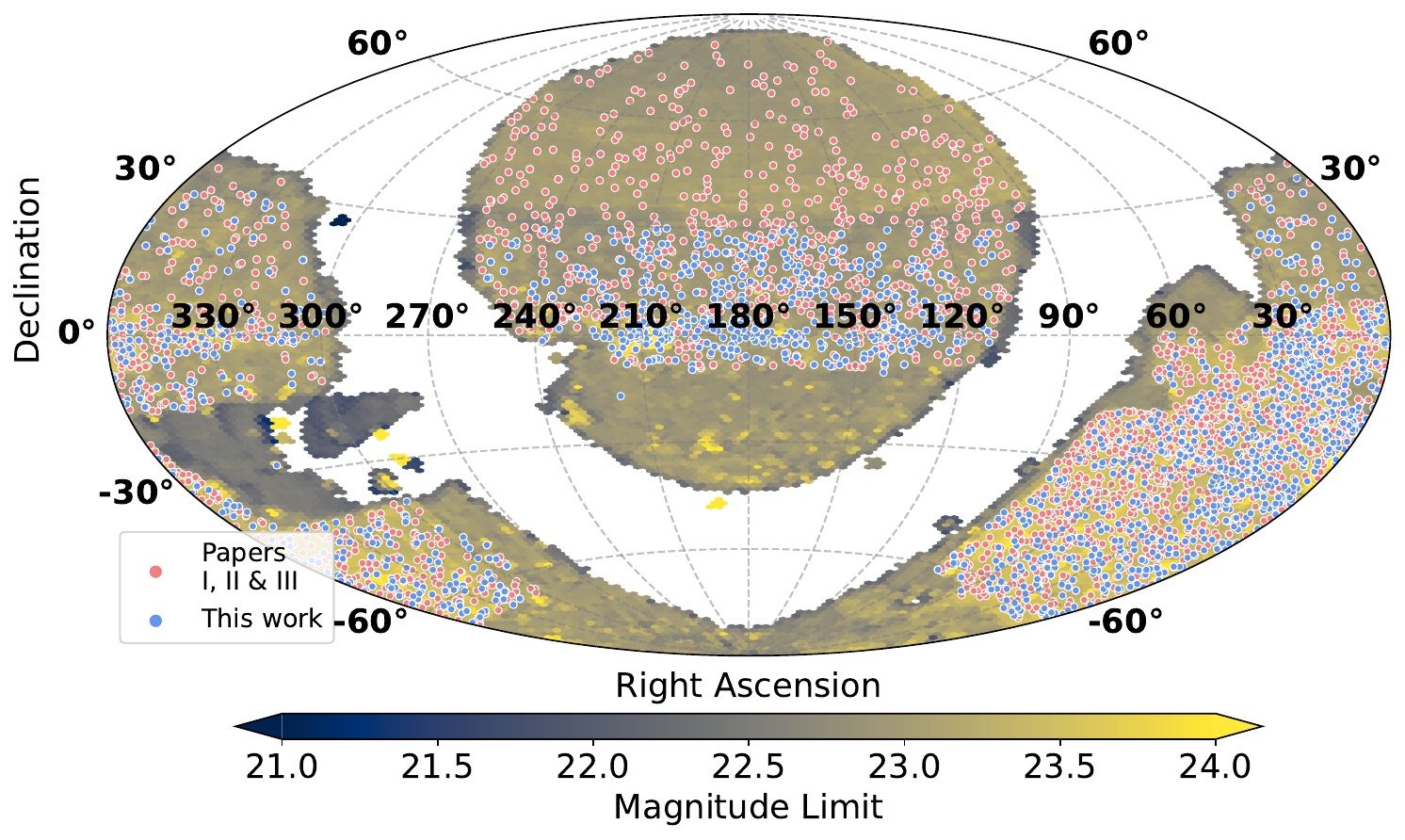}}   
\captionof{figure}{\footnotesize All \grandtotnew candidate lensing systems discovered in the Legacy Surveys reported in Papers~I through III (red), and this work (blue), over the hexagonal bin depth map of the Legacy Surveys DR10, including the footprint of the BASS/MzLS surveys.}
\label{fig:cand_map_all}
\end{minipage}

\vspace{0.5cm}
\begin{minipage}{\linewidth}
\makebox[\linewidth]{
  \includegraphics[keepaspectratio=true,scale=0.7]{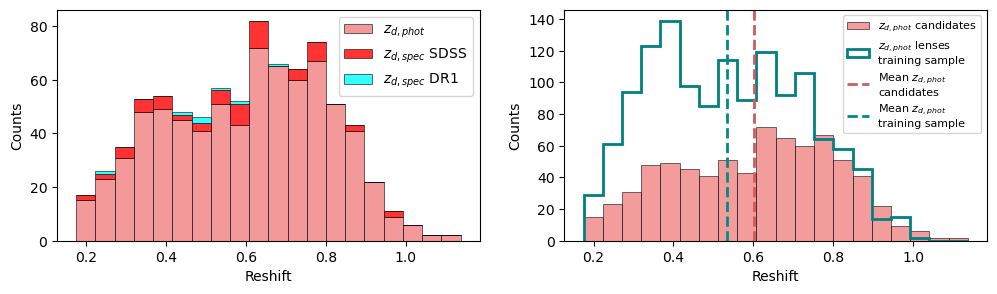}} 
\captionof{figure}{\footnotesize \textbf{Left}: lens spectroscopic redshift, $z_{\text{d,spec}}$, stacked distributions of the candidates found in this work (SDSS Data Release 17 in dark red and DESI DR1 \cite{DR1} 
in cyan), and photometric redshift, $z_{\text{d,phot}}$, distribution from \cite{zhou2021a} in light red. When DR1 redshift is not available, SDSS redshift is reported. In the absence of DR1 and SDSS redshifts, photometric redshift is provided.
A DR1 redshift outlier of $z=1.669$ corresponding to candidate DESI-006.2804+03.3820 was omitted from this histogram because visual inspection of its spectrum revealed no spectroscopic features to support the 
\emph{Redrock} (see \url{https://github.com/desihub/redrock/releases/tag/0.15.4}) redshift.
\textbf{Right}: lens redshift distributions of the new candidates discovered in this work (light red) and the lenses included in the training sample (teal outline). Dashed vertical lines indicate the average photometric redshift of the training sample and the candidates.}
\label{fig:z-distributions}
\end{minipage}

\newpage
\section{Discussion}\label{sec:discussion}
This search in DR10 employs two neural network architectures: a custom ResNet (\S\ref{sec:shielded}) and an EfficientNet (\S\ref{sec:efficientnet}). Table~\ref{tab:model-comparison} compares the sizes, training times, and AUCs of these two models. Notably, both trained in approximately the same amount of time ($\sim1$ hour), despite the EfficientNet having $\sim$100 times more trainable parameters than ResNet. This stark difference in size, combined with the similar training durations, highlights EfficientNet's scalability. Conversely, our custom ResNet is impressive in that it achieves virtually identical AUC performance as EfficientNet despite having far fewer trainable parameters.
\begin{deluxetable*}{lccccccc}[h]
\tablewidth{0pt}
\tabletypesize{\scriptsize}
\tablecaption{Model size, training time, and AUC comparison.}\label{tab:model-comparison}
\tablehead{\colhead{} & \colhead{EfficientNet} & \colhead{ResNet}}
\startdata
    Trainable Parameters & 20,542,883& 194,433 \\
    Training Time [mins]& 61& 56 \\
    Highest AUC & 0.9987& 0.9984 \\
 \hline
\enddata
\end{deluxetable*}

Figure~\ref{fig:probabilities-comparison} compares the output probabilities assigned by the ResNet and the EfficientNet for new candidates graded A, B, and C. Although both models achieve similar AUCs, they exhibit slight differences in their probability distributions. The EfficientNet assigns a broader range of probabilities, extending down to $\sim0.86$, whereas the ResNet probabilities only drop to $\sim0.92$. This suggests that the ResNet is more confident in its predictions, while the EfficientNet is more conservative.
Despite these differences, the marginal histograms show that both models concentrate candidates in the upper probability range, with all grades peaking at the highest probability bin.

\vspace{0.5cm}
\begin{minipage}{\linewidth}
\makebox[\linewidth]{
  \includegraphics[keepaspectratio=true,scale=0.55]{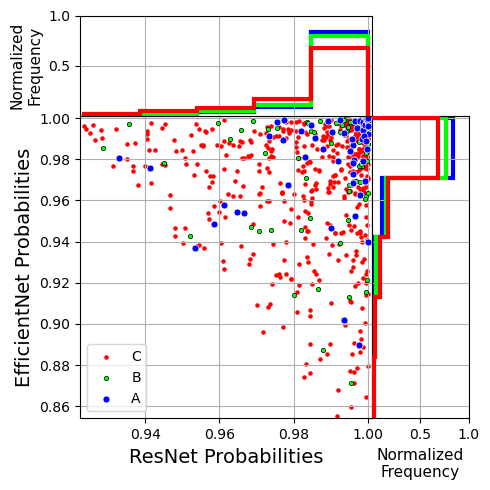}}
\captionof{figure}{\footnotesize Comparison of the ResNet and EfficientNet output probabilities for candidates by grade. Each point represents a candidate, with its position indicating the probability assigned by ResNet ($x$-axis) and EfficientNet ($y$-axis). The marginal histograms show the probability distributions for each grade, normalized such that the total frequency for each \textit{grade} sums to 1.}
\label{fig:probabilities-comparison}
\end{minipage}

While other lens searches, such as \cite{shu2022a}, have implemented two different classifiers simultaneously, we advance this technique by using a meta-learner to aggregate the probabilities of our base models (see \S\ref{sec:meta-learner}). This method of combining probabilities results in an overall AUC higher than that of the individual models, as shown in Figure~\ref{fig:auc-comparison}. 
While in this case, the performance of the meta-learner is the same as the simple average of the predictions from the two models, this nevertheless provides a systematic approach to optimally combine the results from multiple models in future lens searches.

The use of individual \textit{grz} bands for visual inspection (see \S\ref{sec:VI-portal}) has proven to be highly effective. As demonstrated in Figure~\ref{fig:VI_demo}, different bands can help distinguish the lens from the source. This capability is particularly valuable in cases where the Einstein radius is comparable to the seeing.

% \newpage
\section{Conclusions}\label{sec:conclusion}
We have carried out a search for strong gravitational lensing systems in the DESI Legacy Surveys Data Release 10 (DR10). This is the fourth paper in a series of lens searches in the DESI Legacy Surveys, following Paper~I \citep[][DR7 DECaLS, {\tt DEV} and {\tt COMP} only]{huang2020a}, Paper~II \citep[][DR8, {\tt DEV}, {\tt COMP}, and {\tt REX}]{huang2021a}, and Paper~III \citep[][DR9, {\tt SER}, {\tt DEV}, {\tt REX}, and {\tt EXP}]{storfer2024a}.
We use a neural network ensemble consisting of a custom ResNet (\citealp{lanusse2018a}; Paper~II) 
and an EfficientNet \citep{Tan2019},
trained on observed lenses and nonlenses.
We apply our trained models to $\sim$~43 million non-{\tt PSF} (i.e., {\tt SER}, {\tt DEV}, {\tt REX}, and {\tt EXP}, extracted and typed by \tractor ) 
cutout images with at least three exposures in each of the $grz$ bands and a $z$-band magnitude cut of $< 20.0$~mag for the galaxy at the center of each image.

We hold a high standard in grading these candidate systems.
We report \NnewlensesUndisc \textit{new} lens candidates with a grade breakdown of \NnewlensesUndiscA As, \NnewlensesUndiscB Bs, and \NnewlensesUndiscC Cs.
Combining all four searches, we have found a grand total of new \Nnewlensesacrosspapers strong lens candidates. 
Grade~D systems are not included in the counts, but they can be found on our project website (URL provided in \S\ref{sec:all-cands}).

In this search we have made improvements in four major areas: machine learning methods, training sample, neural network training hardware, and visual inspection:

\begin{itemize}
    \item We have implemented a neural network ensemble composed of models with two distinct architectures whose predictions are optimally combined using a meta-learner, as opposed to a single architecture (Papers I-III). 
    \item We continue to use observed images for training as in Papers~I--III, but we have increased the size of our training sample by increasing the nonlens-to-lens ratio from 33:1 to 100:1, respectively, from Paper III to this work. The total training sample size thus has increased from 66,545 to 135,554 between Paper~III and this work.
    \item We have trained both models on four NVIDIA A100 GPUs on a single GPU node, whereas for Papers I-III, the training was done on one GPU. 
    Using 4 GPUs results in approximately 3.7~and~$3\times$ speed gain for the ResNet and EfficientNet, respectively.
    Our training and hyperparameter optimization for two models with a training sample that is twice as large as for Paper~III
    takes advantage of the speed afforded by the parallelized training.  
    \item  We use both the combined RGB and individual \textit{grz} band images during visual inspection. This makes it easier to identify systems that have small Einstein radius and/or faint lensed arc(s), but sufficient color difference between the lens and lensed arcs.
    This work also uses a specialized portal that streamlines the process by enabling the inspection and annotation of candidates efficiently. 
\end{itemize}

With the improvements and innovations outlined above, we find an increase in purity in this work where 1 in every \SERpurity~{\tt SER}, \DEVpurity~{\tt DEV}, \REXpurity~{\tt REX}, and \EXPpurity~{\tt EXP} neural network recommendations is a lens candidate. 

To our knowledge, the additional footprint covered by DR10 compared to DR9 (see Figure~\ref{fig:dr9-footprint}) has not been searched for lenses. In this sub-region, we have found 327 lens candidates.
However, in all other parts of the DECaLS footprint, multiple lens searches have been performed, by our group or others. And yet, we have found 484 \emph{new} candidates in these parts. 
Therefore, systematic comparisons of lens searches to investigate the 
demographics (the ``selection function'') of lens candidates (especially those with A and B grades) found and missed by various searches covering the same footprint would be beneficial to astrophysics and machine learning. 
Work in this direction has begun \citep[e.g.,][]{knabel2020,more2024} and the search results from this work make it clear that more such studies are needed.

This work, together with other searches, has clearly demonstrated that machine learning approaches are highly effective in discovering large numbers of high-quality strong lensing candidates. This will certainly continue to be the case for future surveys such as the Vera C. Rubin Observatory Legacy Survey of Space and Time (LSST), Euclid, and the Nancy Grace Roman Space Telescope. The discovery of strong lenses will therefore continue to accelerate 
and the potential for using strong lensing to tackle major astrophysical and cosmological questions has never been greater.

\newpage
\section{Acknowledgement}\label{sec:acknowledgement}
This research used resources of the National Energy Research Scientific Computing Center (NERSC), a U.S. Department of Energy Office of Science User Facility operated under Contract No. DE-AC02-05CH11231 and the Computational HEP program in The Department of Energy’s Science Office of High Energy Physics provided resources through the “Cosmology Data Repository” project (Grant \#KA2401022)
X.H. acknowledges the University of San Francisco Faculty Development Fund. 
A.D.'s research is supported by National Science Foundation's National Optical-Infrared Astronomy Research Laboratory, 
which is operated by the Association of Universities for Research in Astronomy (AURA) under cooperative agreement with the National Science Foundation.

This paper is based on observations at Cerro Tololo Inter-American Observatory, National Optical
Astronomy Observatory (NOAO Prop. ID: 2014B-0404; co-PIs: D. J. Schlegel and A. Dey), which is operated by the Association of
Universities for Research in Astronomy (AURA) under a cooperative agreement with the
National Science Foundation.

This project used data obtained with the Dark Energy Camera (DECam),
which was constructed by the Dark Energy Survey (DES) collaboration.
Funding for the DES Projects has been provided by 
the U.S. Department of Energy, 
the U.S. National Science Foundation, 
the Ministry of Science and Education of Spain, 
the Science and Technology Facilities Council of the United Kingdom, 
the Higher Education Funding Council for England, 
the National Center for Supercomputing Applications at the University of Illinois at Urbana-Champaign, 
the Kavli Institute of Cosmological Physics at the University of Chicago, 
the Center for Cosmology and Astro-Particle Physics at The Ohio State University, 
the Mitchell Institute for Fundamental Physics and Astronomy at Texas A\&M University, 
Financiadora de Estudos e Projetos, Funda{\c c}{\~a}o Carlos Chagas Filho de Amparo {\`a} Pesquisa do Estado do Rio de Janeiro, 
Conselho Nacional de Desenvolvimento Cient{\'i}fico e Tecnol{\'o}gico and the Minist{\'e}rio da Ci{\^e}ncia, Tecnologia e Inovac{\~a}o, 
the Deutsche Forschungsgemeinschaft, 
and the Collaborating Institutions in the Dark Energy Survey.
The Collaborating Institutions are 
Argonne National Laboratory, 
the University of California at Santa Cruz, 
the University of Cambridge, 
Centro de Investigaciones En{\'e}rgeticas, Medioambientales y Tecnol{\'o}gicas-Madrid, 
the University of Chicago, 
University College London, 
the DES-Brazil Consortium, 
the University of Edinburgh, 
the Eidgen{\"o}ssische Technische Hoch\-schule (ETH) Z{\"u}rich, 
Fermi National Accelerator Laboratory, 
the University of Illinois at Urbana-Champaign, 
the Institut de Ci{\`e}ncies de l'Espai (IEEC/CSIC), 
the Institut de F{\'i}sica d'Altes Energies, 
Lawrence Berkeley National Laboratory, 
the Ludwig-Maximilians Universit{\"a}t M{\"u}nchen and the associated Excellence Cluster Universe, 
the University of Michigan, 
{the} National Optical Astronomy Observatory, 
the University of Nottingham, 
The Ohio State University, 
the OzDES Membership Consortium
the University of Pennsylvania, 
the University of Portsmouth, 
SLAC National Accelerator Laboratory, 
Stanford University, 
the University of Sussex, 
and Texas A\&M University.

\bibliography{dustarchive}

\begin{thebibliography}{}
\expandafter\ifx\csname natexlab\endcsname\relax\def\natexlab#1{#1}\fi
\providecommand{\url}[1]{\href{#1}{#1}}
\providecommand{\dodoi}[1]{doi:~\href{http://doi.org/#1}{\nolinkurl{#1}}}
\providecommand{\doeprint}[1]{\href{http://ascl.net/#1}{\nolinkurl{http://ascl.net/#1}}}
\providecommand{\doarXiv}[1]{\href{https://arxiv.org/abs/#1}{\nolinkurl{https://arxiv.org/abs/#1}}}

\bibitem[{{Anguita} {et~al.}(2012){Anguita}, {Barrientos}, {Gladders}, {Faure}, {Yee}, \& {Gilbank}}]{anguita2012}
{Anguita}, T., {Barrientos}, L.~F., {Gladders}, M.~D., {et~al.} 2012, \apj, 748, 129, \dodoi{10.1088/0004-637X/748/2/129}

\bibitem[{{Birrer} {et~al.}(2020){Birrer}, {Shajib}, {Galan}, {Millon}, {Treu}, {Agnello}, {Auger}, {Chen}, {Christensen}, {Collett}, {Courbin}, {Fassnacht}, {Koopmans}, {Marshall}, {Park}, {Rusu}, {Sluse}, {Spiniello}, {Suyu}, {Wagner-Carena}, {Wong}, {Barnab{\`e}}, {Bolton}, {Czoske}, {Ding}, {Frieman}, \& {Van de Vyvere}}]{birrer2020a}
{Birrer}, S., {Shajib}, A.~J., {Galan}, A., {et~al.} 2020, arXiv e-prints, arXiv:2007.02941.
\newblock \doarXiv{2007.02941}

\bibitem[{{Bolton} {et~al.}(2006){Bolton}, {Burles}, {Koopmans}, {Treu}, \& {Moustakas}}]{bolton2006a}
{Bolton}, A.~S., {Burles}, S., {Koopmans}, L.~V.~E., {Treu}, T., \& {Moustakas}, L.~A. 2006, \apj, 638, 703, \dodoi{10.1086/498884}

\bibitem[{{Broadhurst} {et~al.}(2000){Broadhurst}, {Huang}, {Frye}, \& {Ellis}}]{broadhurst2000a}
{Broadhurst}, T., {Huang}, X., {Frye}, B., \& {Ellis}, R. 2000, \apjl, 534, L15, \dodoi{10.1086/312651}

\bibitem[{{Brownstein} {et~al.}(2012){Brownstein}, {Bolton}, {Schlegel}, {Eisenstein}, {Kochanek}, {Connolly}, {Maraston}, {Pandey}, {Seitz}, {Wake}, {Wood-Vasey}, {Brinkmann}, {Schneider}, \& {Weaver}}]{brownstein2012a}
{Brownstein}, J.~R., {Bolton}, A.~S., {Schlegel}, D.~J., {et~al.} 2012, \apj, 744, 41, \dodoi{10.1088/0004-637X/744/1/41}

\bibitem[{{Ca{\~n}ameras} {et~al.}(2020){Ca{\~n}ameras}, {Schuldt}, {Suyu}, {Taubenberger}, {Meinhardt}, {Leal-Taixe}, {Lemon}, {Rojas}, \& {Savary}}]{canameras2020a}
{Ca{\~n}ameras}, R., {Schuldt}, S., {Suyu}, S.~H., {et~al.} 2020, arXiv e-prints, arXiv:2004.13048.
\newblock \doarXiv{2004.13048}

\bibitem[{{Ca{\~n}ameras} {et~al.}(2021){Ca{\~n}ameras}, {Schuldt}, {Shu}, {Suyu}, {Taubenberger}, {Meinhardt}, {Leal-Taix{\'e}}, {Chao}, {Inoue}, {Jaelani}, \& {More}}]{canameras2021a}
{Ca{\~n}ameras}, R., {Schuldt}, S., {Shu}, Y., {et~al.} 2021, \aap, 653, L6, \dodoi{10.1051/0004-6361/202141758}

\bibitem[{{Caminha} {et~al.}(2022){Caminha}, {Suyu}, {Grillo}, \& {Rosati}}]{caminha2022a}
{Caminha}, G.~B., {Suyu}, S.~H., {Grillo}, C., \& {Rosati}, P. 2022, \aap, 657, A83, \dodoi{10.1051/0004-6361/202141994}

\bibitem[{{Carrasco} {et~al.}(2017){Carrasco}, {Barrientos}, {Anguita}, {Garc{\'{\i}}a-Vergara}, {Bayliss}, {Gladders}, {Gilbank}, {Yee}, \& {West}}]{carrasco2017a}
{Carrasco}, M., {Barrientos}, L.~F., {Anguita}, T., {et~al.} 2017, \apj, 834, 210, \dodoi{10.3847/1538-4357/834/2/210}

\bibitem[{{{\c{C}}a{\u{g}}an {\c{S}}eng{\"u}l} {et~al.}(2021){{\c{C}}a{\u{g}}an {\c{S}}eng{\"u}l}, {Dvorkin}, {Ostdiek}, \& {Tsang}}]{cagansengul2021a}
{{\c{C}}a{\u{g}}an {\c{S}}eng{\"u}l}, A., {Dvorkin}, C., {Ostdiek}, B., \& {Tsang}, A. 2021, arXiv e-prints, arXiv:2112.00749.
\newblock \doarXiv{2112.00749}

\bibitem[{{Chan} {et~al.}(2020){Chan}, {Suyu}, {Sonnenfeld}, {Jaelani}, {More}, {Yonehara}, {Kubota}, {Coupon}, {Lee}, {Oguri}, {Rusu}, \& {Wong}}]{chan2020a}
{Chan}, J. H.~H., {Suyu}, S.~H., {Sonnenfeld}, A., {et~al.} 2020, \aap, 636, A87, \dodoi{10.1051/0004-6361/201937030}

\bibitem[{{Chen} {et~al.}(2022){Chen}, {Kelly}, {Oguri}, {Broadhurst}, {Diego}, {Emami}, {Filippenko}, {Treu}, \& {Zitrin}}]{chen2022a}
{Chen}, W., {Kelly}, P.~L., {Oguri}, M., {et~al.} 2022, \nat, 611, 256, \dodoi{10.1038/s41586-022-05252-5}

\bibitem[{{Collett} \& {Auger}(2014)}]{collett2014a}
{Collett}, T.~E., \& {Auger}, M.~W. 2014, \mnras, 443, 969, \dodoi{10.1093/mnras/stu1190}

\bibitem[{Coscrato {et~al.}(2020)Coscrato, de~Almeida~In{\'a}cio, \& Izbicki}]{COSCRATO2020141}
Coscrato, V., de~Almeida~In{\'a}cio, M.~H., \& Izbicki, R. 2020, Neurocomputing, 399, 141, \dodoi{https://doi.org/10.1016/j.neucom.2020.02.073}

\bibitem[{{Dark Energy Survey Collaboration} {et~al.}(2016){Dark Energy Survey Collaboration}, {Abbott}, {Abdalla}, {Aleksi{\'c}}, {Allam}, {Amara}, {Bacon}, {Balbinot}, {Banerji}, {Bechtol}, {Benoit-L{\'e}vy}, {Bernstein}, {Bertin}, {Blazek}, {Bonnett}, {Bridle}, {Brooks}, {Brunner}, {Buckley-Geer}, {Burke}, {Caminha}, {Capozzi}, {Carlsen}, {Carnero-Rosell}, {Carollo}, {Carrasco-Kind}, {Carretero}, {Castander}, {Clerkin}, {Collett}, {Conselice}, {Crocce}, {Cunha}, {D'Andrea}, {da Costa}, {Davis}, {Desai}, {Diehl}, {Dietrich}, {Dodelson}, {Doel}, {Drlica-Wagner}, {Estrada}, {Etherington}, {Evrard}, {Fabbri}, {Finley}, {Flaugher}, {Foley}, {Fosalba}, {Frieman}, {Garc{\'\i}a-Bellido}, {Gaztanaga}, {Gerdes}, {Giannantonio}, {Goldstein}, {Gruen}, {Gruendl}, {Guarnieri}, {Gutierrez}, {Hartley}, {Honscheid}, {Jain}, {James}, {Jeltema}, {Jouvel}, {Kessler}, {King}, {Kirk}, {Kron}, {Kuehn}, {Kuropatkin}, {Lahav}, {Li}, {Lima}, {Lin}, {Maia}, {Makler}, {Manera}, {Maraston}, {Marshall}, {Martini}, {McMahon},
  {Melchior}, {Merson}, {Miller}, {Miquel}, {Mohr}, {Morice-Atkinson}, {Naidoo}, {Neilsen}, {Nichol}, {Nord}, {Ogando}, {Ostrovski}, {Palmese}, {Papadopoulos}, {Peiris}, {Peoples}, {Percival}, {Plazas}, {Reed}, {Refregier}, {Romer}, {Roodman}, {Ross}, {Rozo}, {Rykoff}, {Sadeh}, {Sako}, {S{\'a}nchez}, {Sanchez}, {Santiago}, {Scarpine}, {Schubnell}, {Sevilla-Noarbe}, {Sheldon}, {Smith}, {Smith}, {Soares-Santos}, {Sobreira}, {Soumagnac}, {Suchyta}, {Sullivan}, {Swanson}, {Tarle}, {Thaler}, {Thomas}, {Thomas}, {Tucker}, {Vieira}, {Vikram}, {Walker}, {Wechsler}, {Weller}, {Wester}, {Whiteway}, {Wilcox}, {Yanny}, {Zhang}, \& {Zuntz}}]{des2016}
{Dark Energy Survey Collaboration}, {Abbott}, T., {Abdalla}, F.~B., {et~al.} 2016, \mnras, 460, 1270, \dodoi{10.1093/mnras/stw641}

\bibitem[{{Dawes} {et~al.}(2023){Dawes}, {Storfer}, {Huang}, {Aldering}, {Cikota}, {Dey}, \& {Schlegel}}]{dawes2023}
{Dawes}, C., {Storfer}, C., {Huang}, X., {et~al.} 2023, \apjs, 269, 61, \dodoi{10.3847/1538-4365/ad015a}

\bibitem[{{DESI Collaboration} {et~al.}(2025){DESI Collaboration}, {Abdul-Karim}, {Adame}, {Aguado}, {Aguilar}, {Ahlen}, {Alam}, {Aldering}, {Alexander}, {Alfarsy}, {Allen}, {Allende Prieto}, {Alves}, {Anand}, {Andrade}, {Armengaud}, {Avila}, {Aviles}, {Awan}, {Bailey}, {Baleato Lizancos}, {Ballester}, {Bault}, {Bautista}, {BenZvi}, {Beraldo e Silva}, {Bermejo-Climent}, {Beutler}, {Bianchi}, {Blake}, {Blum}, {Bolton}, {Bonici}, {Brieden}, {Brodzeller}, {Brooks}, {Buckley-Geer}, {Burtin}, {Canning}, {Carnero Rosell}, {Carr}, {Carrilho}, {Casas}, {Castander}, {Cereskaite}, {Cervantes-Cota}, {Chaussidon}, {Chaves-Montero}, {Chen}, {Chen}, {Claybaugh}, {Cole}, {Cooper}, {Cousinou}, {Cuceu}, {Davis}, {Dawson}, {de Belsunce}, {de la Cruz}, {de la Macorra}, {de Mattia}, {Deiosso}, {Della Costa}, {Demina}, {Demirbozan}, {DeRose}, {Dey}, {Dey}, {Ding}, {Ding}, {Doel}, {Douglass}, {Dowicz}, {Ebina}, {Edelstein}, {Eisenstein}, {Elbers}, {Emas}, {Escoffier}, {Fagrelius}, {Fan}, {Fanning}, {Fawcett},
  {Fern{\'a}ndez-Garc{\'i}a}, {Ferraro}, {Findlay}, {Font-Ribera}, {Forero-Romero}, {Forero-S{\'a}nchez}, {Frenk}, {G{\"a}nsicke}, {Galbany}, {Garc{\'i}a-Bellido}, {Garcia-Quintero}, {Garrison}, {Gazta{\~n}aga}, {Gil-Mar{\'i}n}, {Gnedin}, {Gontcho}, {Gonzalez-Morales}, {Gonzalez-Perez}, {Gordon}, {Graur}, {Green}, {Gruen}, {Gsponer}, {Guandalin}, {Gutierrez}, {Guy}, {Hahn}, {Han}, {Han}, {He}, {Herrera-Alcantar}, {Honscheid}, {Hou}, {Howlett}, {Huterer}, {Ir\v{s}i\v{c}}, {Ishak}, {Jacques}, {Jimenez}, {Jing}, {Joachimi}, {Joudaki}, {Joyce}, {Jullo}, {Juneau}, {Kara\c{c}ayl{\i}}, {Karim}, {Kehoe}, {Kent}, {Khederlarian}, {Kirkby}, {Kisner}, {Kitaura}, {Kizhuprakkat}, {Kong}, {Koposov}, {Kremin}, {Krolewski}, {Lahav}, {Lai}, {Lamman}, {Lan}, {Landriau}, {Lang}, {Lange}, {Lasker}, {Le Goff}, {Le Guillou}, {Leauthaud}, {Levi}, {Li}, {Li}, {Lodha}, {Lokken}, {Luo}, {Magneville}, {Manera}, {Manser}, {Margala}, {Martini}, {Maus}, {McCullough}, {McDonald}, {Medina}, {Medina-Varela}, {Meisner}, {Mena-Fern{\'a}ndez},
  {Menegas}, {Mezcua}, {Miquel}, {Montero-Camacho}, {Moon}, {Moustakas}, {Mu{\~n}oz-Guti{\'e}rrez}, {Mu{\~n}oz-Santos}, {Myers}, {Myles}, {Nadathur}, {Najita}, {Napolitano}, {Newman}, {Nikakhtar}, {Nikutta}, {Niz}, {Noriega}, {Padmanabhan}, {Paillas}, {Palanque-Delabrouille}, {Palmese}, {Pan}, {Pan}, {Parkinson}, {Peacock}, {Percival}, {P{\'e}rez-Fern{\'a}ndez}, {P{\'e}rez-R\`afols}, \& {Peterson}}]{DR1}
{DESI Collaboration}, {Abdul-Karim}, M., {Adame}, A.~G., {et~al.} 2025, arXiv e-prints, arXiv:2503.14745, \dodoi{10.48550/arXiv.2503.14745}

\bibitem[{Dey {et~al.}(2019)Dey, Schlegel, Lang, Blum, Burleigh, Fan, Findlay, Finkbeiner, Herrer, Juneau, Knezek, Leisenring, Levi, Mack, Meisner, Myers, Newman, Schlafly, Shectman, Stupak, Valdes, Walker, Weaver, Weinberg, \& Zhao}]{Dey2019}
Dey, A., Schlegel, D.~J., Lang, D., {et~al.} 2019, The Astronomical Journal, 157, 168, \dodoi{10.3847/1538-3881/ab089d}

\bibitem[{{Diehl} {et~al.}(2017){Diehl}, {Buckley-Geer}, {Lindgren}, {Nord}, {Gaitsch}, {Gaitsch}, {Lin}, {Allam}, {Collett}, {Furlanetto}, {Gill}, {More}, {Nightingale}, {Odden}, {Pellico}, {Tucker}, {da Costa}, {Fausti Neto}, {Kuropatkin}, {Soares-Santos}, {Welch}, {Zhang}, {Frieman}, {Abdalla}, {Annis}, {Benoit-L{\'e}vy}, {Bertin}, {Brooks}, {Burke}, {Carnero Rosell}, {Carrasco Kind}, {Carretero}, {Cunha}, {D'Andrea}, {Desai}, {Dietrich}, {Drlica-Wagner}, {Evrard}, {Finley}, {Flaugher}, {Garc{\'{\i}}a-Bellido}, {Gerdes}, {Goldstein}, {Gruen}, {Gruendl}, {Gschwend}, {Gutierrez}, {James}, {Kuehn}, {Kuhlmann}, {Lahav}, {Li}, {Lima}, {Maia}, {Marshall}, {Menanteau}, {Miquel}, {Nichol}, {Nugent}, {Ogando}, {Plazas}, {Reil}, {Romer}, {Sako}, {Sanchez}, {Santiago}, {Scarpine}, {Schindler}, {Schubnell}, {Sevilla-Noarbe}, {Sheldon}, {Smith}, {Sobreira}, {Suchyta}, {Swanson}, {Tarle}, {Thomas}, {Walker}, \& {DES Collaboration}}]{diehl2017a}
{Diehl}, H.~T., {Buckley-Geer}, E.~J., {Lindgren}, K.~A., {et~al.} 2017, \apjs, 232, 15, \dodoi{10.3847/1538-4365/aa8667}

\bibitem[{{Ding} {et~al.}(2021){Ding}, {Liao}, {Birrer}, {Shajib}, {Treu}, \& {Yang}}]{ding2021a}
{Ding}, X., {Liao}, K., {Birrer}, S., {et~al.} 2021, \mnras, 504, 5621, \dodoi{10.1093/mnras/stab1240}

\bibitem[{{Drlica-Wagner} {et~al.}(2021){Drlica-Wagner}, {Carlin}, {Nidever}, {Ferguson}, {Kuropatkin}, {Adam{\'o}w}, {Cerny}, {Choi}, {Esteves}, {Mart{\'\i}nez-V{\'a}zquez}, {Mau}, {Miller}, {Mutlu-Pakdil}, {Neilsen}, {Olsen}, {Pace}, {Riley}, {Sakowska}, {Sand}, {Santana-Silva}, {Tollerud}, {Tucker}, {Vivas}, {Zaborowski}, {Zenteno}, {Abbott}, {Allam}, {Bechtol}, {Bell}, {Bell}, {Bilaji}, {Bom}, {Carballo-Bello}, {Crnojevi{\'c}}, {Cioni}, {Diaz-Ocampo}, {de Boer}, {Erkal}, {Gruendl}, {Hernandez-Lang}, {Hughes}, {James}, {Johnson}, {Li}, {Mao}, {Mart{\'\i}nez-Delgado}, {Massana}, {McNanna}, {Morgan}, {Nadler}, {No{\"e}l}, {Palmese}, {Peter}, {Rykoff}, {S{\'a}nchez}, {Shipp}, {Simon}, {Smercina}, {Soares-Santos}, {Stringfellow}, {Tavangar}, {van der Marel}, {Walker}, {Wechsler}, {Wu}, {Yanny}, {Fitzpatrick}, {Huang}, {Jacques}, {Nikutta}, {Scott}, \& {Astro Data Lab}}]{Drlica-Wagner2021}
{Drlica-Wagner}, A., {Carlin}, J.~L., {Nidever}, D.~L., {et~al.} 2021, \apjs, 256, 2, \dodoi{10.3847/1538-4365/ac079d}

\bibitem[{{Fagin} {et~al.}(2024){Fagin}, {Vernardos}, {Tsagkatakis}, {Pantazis}, {Shajib}, \& {O'Dowd}}]{fagin2024a}
{Fagin}, J., {Vernardos}, G., {Tsagkatakis}, G., {et~al.} 2024, arXiv e-prints, arXiv:2403.13881, \dodoi{10.48550/arXiv.2403.13881}

\bibitem[{{Flaugher} {et~al.}(2015){Flaugher}, {Diehl}, {Honscheid}, {Abbott}, {Alvarez}, {Angstadt}, {Annis}, {Antonik}, {Ballester}, {Beaufore}, {Bernstein}, {Bernstein}, {Bigelow}, {Bonati}, {Boprie}, {Brooks}, {Buckley-Geer}, {Campa}, {Cardiel-Sas}, {Castander}, {Castilla}, {Cease}, {Cela-Ruiz}, {Chappa}, {Chi}, {Cooper}, {da Costa}, {Dede}, {Derylo}, {DePoy}, {de Vicente}, {Doel}, {Drlica-Wagner}, {Eiting}, {Elliott}, {Emes}, {Estrada}, {Fausti Neto}, {Finley}, {Flores}, {Frieman}, {Gerdes}, {Gladders}, {Gregory}, {Gutierrez}, {Hao}, {Holland}, {Holm}, {Huffman}, {Jackson}, {James}, {Jonas}, {Karcher}, {Karliner}, {Kent}, {Kessler}, {Kozlovsky}, {Kron}, {Kubik}, {Kuehn}, {Kuhlmann}, {Kuk}, {Lahav}, {Lathrop}, {Lee}, {Levi}, {Lewis}, {Li}, {Mandrichenko}, {Marshall}, {Martinez}, {Merritt}, {Miquel}, {Mu{\~n}oz}, {Neilsen}, {Nichol}, {Nord}, {Ogando}, {Olsen}, {Palaio}, {Patton}, {Peoples}, {Plazas}, {Rauch}, {Reil}, {Rheault}, {Roe}, {Rogers}, {Roodman}, {Sanchez}, {Scarpine}, {Schindler}, {Schmidt},
  {Schmitt}, {Schubnell}, {Schultz}, {Schurter}, {Scott}, {Serrano}, {Shaw}, {Smith}, {Soares-Santos}, {Stefanik}, {Stuermer}, {Suchyta}, {Sypniewski}, {Tarle}, {Thaler}, {Tighe}, {Tran}, {Tucker}, {Walker}, {Wang}, {Watson}, {Weaverdyck}, {Wester}, {Woods}, {Yanny}, \& {DES Collaboration}}]{flaugher2015a}
{Flaugher}, B., {Diehl}, H.~T., {Honscheid}, K., {et~al.} 2015, \aj, 150, 150, \dodoi{10.1088/0004-6256/150/5/150}

\bibitem[{{Freedman} {et~al.}(2020){Freedman}, {Madore}, {Hoyt}, {Jang}, {Beaton}, {Lee}, {Monson}, {Neeley}, \& {Rich}}]{freedman2020a}
{Freedman}, W.~L., {Madore}, B.~F., {Hoyt}, T., {et~al.} 2020, \apj, 891, 57, \dodoi{10.3847/1538-4357/ab7339}

\bibitem[{{Gavazzi} {et~al.}(2014){Gavazzi}, {Marshall}, {Treu}, \& {Sonnenfeld}}]{gavazzi2014a}
{Gavazzi}, R., {Marshall}, P.~J., {Treu}, T., \& {Sonnenfeld}, A. 2014, \apj, 785, 144, \dodoi{10.1088/0004-637X/785/2/144}

\bibitem[{{Gonz{\'a}lez} {et~al.}(2025){Gonz{\'a}lez}, {Holloway}, {Collett}, {Verma}, {Bechtol}, {Marshall}, {More}, {Acevedo Barroso}, {Cartwright}, {Martinez}, {Li}, {Rojas}, {Schuldt}, {Birrer}, {Diehl}, {Morgan}, {Drlica-Wagner}, {O'Donnell}, {Zaborowski}, {Nord}, {Baeten}, {Johnson}, {Macmillan}, {Roodman}, {Pieres}, {Walker}, {Plazas Malag{\'o}n}, {Carnero Rosell}, {Santiago}, {Flaugher}, {Gruen}, {Brooks}, {Burke}, {James}, {Sanchez Cid}, {Hollowood}, {Tucker}, {Buckley-Geer}, {Gaztanaga}, {Suchyta}, {Sanchez}, {Gutierrez}, {Giannini}, {Tarle}, {Sevilla-Noarbe}, {Marshall}, {Carretero}, {Frieman}, {De Vicente}, {Garc{\'\i}a-Bellido}, {Mena-Fern{\'a}ndez}, {Myles}, {Honscheid}, {Kuehn}, {Lima}, {Pereira}, {Smith}, {Aguena}, {Weaverdyck}, {Lahav}, {Doel}, {Miquel}, {Gruendl}, {Cawthon}, {Hinton}, {Allam}, {Desai}, {Samuroff}, {Everett}, {Lee}, {Davis}, {Abbott}, \& {Vikram}}]{gonzales2025}
{Gonz{\'a}lez}, J., {Holloway}, P., {Collett}, T., {et~al.} 2025, arXiv e-prints, arXiv:2501.15679, \dodoi{10.48550/arXiv.2501.15679}

\bibitem[{{Grillo} {et~al.}(2015){Grillo}, {Suyu}, {Rosati}, {Mercurio}, {Balestra}, {Munari}, {Nonino}, {Caminha}, {Lombardi}, {De Lucia}, {Borgani}, {Gobat}, {Biviano}, {Girardi}, {Umetsu}, {Coe}, {Koekemoer}, {Postman}, {Zitrin}, {Halkola}, {Broadhurst}, {Sartoris}, {Presotto}, {Annunziatella}, {Maier}, {Fritz}, {Vanzella}, \& {Frye}}]{grillo2015a}
{Grillo}, C., {Suyu}, S.~H., {Rosati}, P., {et~al.} 2015, \apj, 800, 38, \dodoi{10.1088/0004-637X/800/1/38}

\bibitem[{{He} {et~al.}(2023){He}, {Li}, {Cao}, {Li}, {Zou}, \& {Dye}}]{he2023}
{He}, Z., {Li}, N., {Cao}, X., {et~al.} 2023, \aap, 672, A123, \dodoi{10.1051/0004-6361/202245484}

\bibitem[{Hsu(2025)}]{Hsu2025}
Hsu, Y.-M. 2025, spherimatch: Cross-matching and self-matching in spherical coordinates, Astrophysics Source Code Library, ascl:2507.022

\bibitem[{{Huang} {et~al.}(2009){Huang}, {Morokuma}, {Fakhouri}, {Aldering}, {Amanullah}, {Barbary}, {Brodwin}, {Connolly}, {Dawson}, {Doi}, {Faccioli}, {Fadeyev}, {Fruchter}, {Goldhaber}, {Gladders}, {Hennawi}, {Ihara}, {Jee}, {Kowalski}, {Konishi}, {Lidman}, {Meyers}, {Moustakas}, {Perlmutter}, {Rubin}, {Schlegel}, {Spadafora}, {Suzuki}, {Takanashi}, \& {Yasuda}}]{huang2009a}
{Huang}, X., {Morokuma}, T., {Fakhouri}, H.~K., {et~al.} 2009, \apj, 707, L12, \dodoi{10.1088/0004-637X/707/1/L12}

\bibitem[{Huang {et~al.}(2020)Huang, Storfer, Ravi, Pilon, Domingo, Schlegel, Bailey, Dey, Gupta, Herrera, Juneau, Landriau, Lang, Meisner, Moustakas, Myers, Schlafly, Valdes, Weaver, Yang, \& Y{\`{e}}che}]{huang2020a}
Huang, X., Storfer, C., Ravi, V., {et~al.} 2020, The Astrophysical Journal, 894, 78, \dodoi{10.3847/1538-4357/ab7ffb}

\bibitem[{{Huang} {et~al.}(2021){Huang}, {Storfer}, {Gu}, {Ravi}, {Pilon}, {Sheu}, {Venguswamy}, {Banka}, {Dey}, {Landriau}, {Lang}, {Meisner}, {Moustakas}, {Myers}, {Sajith}, {Schlafly}, \& {Schlegel}}]{huang2021a}
{Huang}, X., {Storfer}, C., {Gu}, A., {et~al.} 2021, \apj, 909, 27, \dodoi{10.3847/1538-4357/abd62b}

\bibitem[{{Huber} {et~al.}(2021){Huber}, {Suyu}, {Noebauer}, {Chan}, {Kromer}, {Sim}, {Sluse}, \& {Taubenberger}}]{huber2021a}
{Huber}, S., {Suyu}, S.~H., {Noebauer}, U.~M., {et~al.} 2021, \aap, 646, A110, \dodoi{10.1051/0004-6361/202039218}

\bibitem[{{Jacobs} {et~al.}(2017){Jacobs}, {Glazebrook}, {Collett}, {More}, \& {McCarthy}}]{jacobs2017a}
{Jacobs}, C., {Glazebrook}, K., {Collett}, T., {More}, A., \& {McCarthy}, C. 2017, \mnras, 471, 167, \dodoi{10.1093/mnras/stx1492}

\bibitem[{{Jacobs} {et~al.}(2019{\natexlab{a}}){Jacobs}, {Collett}, {Glazebrook}, {McCarthy}, {Qin}, {Abbott}, {Abdalla}, {Annis}, {Avila}, {Bechtol}, {Bertin}, {Brooks}, {Buckley-Geer}, {Burke}, {Carnero Rosell}, {Carrasco Kind}, {Carretero}, {da Costa}, {Davis}, {De Vicente}, {Desai}, {Diehl}, {Doel}, {Eifler}, {Flaugher}, {Frieman}, {Garc{\'{\i}}a-Bellido}, {Gaztanaga}, {Gerdes}, {Goldstein}, {Gruen}, {Gruendl}, {Gschwend}, {Gutierrez}, {Hartley}, {Hollowood}, {Honscheid}, {Hoyle}, {James}, {Kuehn}, {Kuropatkin}, {Lahav}, {Li}, {Lima}, {Lin}, {Maia}, {Martini}, {Miller}, {Miquel}, {Nord}, {Plazas}, {Sanchez}, {Scarpine}, {Schubnell}, {Serrano}, {Sevilla-Noarbe}, {Smith}, {Soares-Santos}, {Sobreira}, {Suchyta}, {Swanson}, {Tarle}, {Vikram}, {Walker}, {Zhang}, \& {Zuntz}}]{jacobs2019a}
{Jacobs}, C., {Collett}, T., {Glazebrook}, K., {et~al.} 2019{\natexlab{a}}, \mnras, 484, 5330, \dodoi{10.1093/mnras/stz272}

\bibitem[{{Jacobs} {et~al.}(2019{\natexlab{b}}){Jacobs}, {Collett}, {Glazebrook}, {Buckley-Geer}, {Diehl}, {Lin}, {McCarthy}, {Qin}, {Odden}, {Caso Escudero}, {Dial}, {Yung}, {Gaitsch}, {Pellico}, {Lindgren}, {Abbott}, {Annis}, {Avila}, {Brooks}, {Burke}, {Carnero Rosell}, {Carrasco Kind}, {Carretero}, {da Costa}, {De Vicente}, {Fosalba}, {Frieman}, {Garc{\'\i}a-Bellido}, {Gaztanaga}, {Goldstein}, {Gruen}, {Gruendl}, {Gschwend}, {Hollowood}, {Honscheid}, {Hoyle}, {James}, {Krause}, {Kuropatkin}, {Lahav}, {Lima}, {Maia}, {Marshall}, {Miquel}, {Plazas}, {Roodman}, {Sanchez}, {Scarpine}, {Serrano}, {Sevilla-Noarbe}, {Smith}, {Sobreira}, {Suchyta}, {Swanson}, {Tarle}, {Vikram}, {Walker}, {Zhang}, \& {DES Collaboration}}]{jacobs2019b}
---. 2019{\natexlab{b}}, \apjs, 243, 17, \dodoi{10.3847/1538-4365/ab26b6}

\bibitem[{{Jaelani} {et~al.}(2020){Jaelani}, {More}, {Oguri}, {Sonnenfeld}, {Suyu}, {Rusu}, {Wong}, {Chan}, {Kayo}, {Lee}, {Chao}, {Coupon}, {Inoue}, \& {Futamase}}]{jaelani2020a}
{Jaelani}, A.~T., {More}, A., {Oguri}, M., {et~al.} 2020, \mnras, 495, 1291, \dodoi{10.1093/mnras/staa1062}

\bibitem[{Jaelani {et~al.}(2021)Jaelani, Rusu, Kayo, More, Sonnenfeld, Silverman, Schramm, Anguita, Inada, Kondo, Schechter, Lee, Oguri, Chan, Wong, \& Inoue}]{jaelani2021}
Jaelani, A.~T., Rusu, C.~E., Kayo, I., {et~al.} 2021, Monthly Notices of the Royal Astronomical Society, 502, 1487, \dodoi{10.1093/mnras/stab145}

\bibitem[{{Johnson} {et~al.}(2018){Johnson}, {Irwin}, {White}, {Wong}, {Maksym}, {Dupke}, {Miller}, \& {Carrasco}}]{johnson2018}
{Johnson}, L.~E., {Irwin}, J.~A., {White}, III, R.~E., {et~al.} 2018, \apj, 856, 131, \dodoi{10.3847/1538-4357/aab430}

\bibitem[{{Jullo} {et~al.}(2010){Jullo}, {Natarajan}, {Kneib}, {D'Aloisio}, {Limousin}, {Richard}, \& {Schimd}}]{jullo2010a}
{Jullo}, E., {Natarajan}, P., {Kneib}, J.~P., {et~al.} 2010, Science, 329, 924, \dodoi{10.1126/science.1185759}

\bibitem[{{Knabel} {et~al.}(2020){Knabel}, {Steele}, {Holwerda}, {Bridge}, {Jacques}, {Hopkins}, {Bamford}, {Brown}, {Brough}, {Kelvin}, {Bilicki}, \& {Kielkopf}}]{knabel2020}
{Knabel}, S., {Steele}, R.~L., {Holwerda}, B.~W., {et~al.} 2020, \aj, 160, 223, \dodoi{10.3847/1538-3881/abb612}

\bibitem[{{Kochanek}(1991)}]{kochanek1991a}
{Kochanek}, C.~S. 1991, \apj, 373, 354, \dodoi{10.1086/170057}

\bibitem[{{Koopmans} \& {Treu}(2002)}]{koopmans2002a}
{Koopmans}, L.~V.~E., \& {Treu}, T. 2002, \apjl, 568, L5, \dodoi{10.1086/340143}

\bibitem[{{Koopmans} {et~al.}(2006){Koopmans}, {Treu}, {Bolton}, {Burles}, \& {Moustakas}}]{koopmans2006a}
{Koopmans}, L.~V.~E., {Treu}, T., {Bolton}, A.~S., {Burles}, S., \& {Moustakas}, L.~A. 2006, \apj, 649, 599, \dodoi{10.1086/505696}

\bibitem[{{Lang} {et~al.}(2016){Lang}, {Hogg}, \& {Mykytyn}}]{lang2016a}
{Lang}, D., {Hogg}, D.~W., \& {Mykytyn}, D. 2016, {The Tractor: Probabilistic astronomical source detection and measurement}, Astrophysics Source Code Library.
\newblock \doeprint{1604.008}

\bibitem[{{Lanusse} {et~al.}(2018){Lanusse}, {Ma}, {Li}, {Collett}, {Li}, {Ravanbakhsh}, {Mandelbaum}, \& {P{\'o}czos}}]{lanusse2018a}
{Lanusse}, F., {Ma}, Q., {Li}, N., {et~al.} 2018, \mnras, 473, 3895, \dodoi{10.1093/mnras/stx1665}

\bibitem[{{Li} {et~al.}(2021){Li}, {Napolitano}, {Spiniello}, {Tortora}, {Kuijken}, {Koopmans}, {Schneider}, {Getman}, {Xie}, {Long}, {Shu}, {Vernardos}, {Huang}, {Covone}, {Dvornik}, {Heymans}, {Hildebrandt}, {Radovich}, \& {Wright}}]{li2021a}
{Li}, R., {Napolitano}, N.~R., {Spiniello}, C., {et~al.} 2021, \apj, 923, 16, \dodoi{10.3847/1538-4357/ac2df0}

\bibitem[{{Li} {et~al.}(2024){Li}, {Collett}, {Krawczyk}, \& {Enzi}}]{li2024a}
{Li}, T., {Collett}, T.~E., {Krawczyk}, C.~M., \& {Enzi}, W. 2024, \mnras, 527, 5311, \dodoi{10.1093/mnras/stad3514}

\bibitem[{{Linder}(2016)}]{linder2016}
{Linder}, E.~V. 2016, \prd, 94, 083510, \dodoi{10.1103/PhysRevD.94.083510}

\bibitem[{{Massey} {et~al.}(2010){Massey}, {Kitching}, \& {Richard}}]{massey2010a}
{Massey}, R., {Kitching}, T., \& {Richard}, J. 2010, Reports on Progress in Physics, 73, 086901, \dodoi{10.1088/0034-4885/73/8/086901}

\bibitem[{{Meneghetti} {et~al.}(2020){Meneghetti}, {Davoli}, {Bergamini}, {Rosati}, {Natarajan}, {Giocoli}, {Caminha}, {Metcalf}, {Rasia}, {Borgani}, {Calura}, {Grillo}, {Mercurio}, \& {Vanzella}}]{meneghetti2020a}
{Meneghetti}, M., {Davoli}, G., {Bergamini}, P., {et~al.} 2020, Science, 369, 1347, \dodoi{10.1126/science.aax5164}

\bibitem[{{Meneghetti} {et~al.}(2023){Meneghetti}, {Cui}, {Rasia}, {Yepes}, {Acebron}, {Angora}, {Bergamini}, {Borgani}, {Calura}, {Despali}, {Giocoli}, {Granata}, {Grillo}, {Knebe}, {Macci{\`o}}, {Mercurio}, {Moscardini}, {Natarajan}, {Ragagnin}, {Rosati}, \& {Vanzella}}]{meneghetti2023a}
{Meneghetti}, M., {Cui}, W., {Rasia}, E., {et~al.} 2023, \aap, 678, L2, \dodoi{10.1051/0004-6361/202346975}

\bibitem[{{Metcalf} {et~al.}(2018){Metcalf}, {Meneghetti}, {Avestruz}, {Bellagamba}, {Bom}, {Bertin}, {Cabanac}, {Davies}, {Decenci{\`e}re}, {Flamary}, {Gavazzi}, {Geiger}, {Hartley}, {Huertas-Company}, {Jackson}, {Jullo}, {Kneib}, {Koopmans}, {Lanusse}, {Li}, {Ma}, {Makler}, {Li}, {Lightman}, {Enrico Petrillo}, {Serjeant}, {Sch{\"a}fer}, {Sonnenfeld}, {Tagore}, {Tortora}, {Tuccillo}, {Valent{\'\i}n}, {Velasco-Forero}, {Verdoes Kleijn}, \& {Vernardos}}]{metcalf2018a}
{Metcalf}, R.~B., {Meneghetti}, M., {Avestruz}, C., {et~al.} 2018, arXiv e-prints, arXiv:1802.03609.
\newblock \doarXiv{1802.03609}

\bibitem[{{Millon} {et~al.}(2020){Millon}, {Galan}, {Courbin}, {Treu}, {Suyu}, {Ding}, {Birrer}, {Chen}, {Shajib}, {Sluse}, {Wong}, {Agnello}, {Auger}, {Buckley-Geer}, {Chan}, {Collett}, {Fassnacht}, {Hilbert}, {Koopmans}, {Motta}, {Mukherjee}, {Rusu}, {Sonnenfeld}, {Spiniello}, \& {Van de Vyvere}}]{millon2020a}
{Millon}, M., {Galan}, A., {Courbin}, F., {et~al.} 2020, \aap, 639, A101, \dodoi{10.1051/0004-6361/201937351}

\bibitem[{{More} {et~al.}(2012){More}, {Cabanac}, {More}, {Alard}, {Limousin}, {Kneib}, {Gavazzi}, \& {Motta}}]{more2012a}
{More}, A., {Cabanac}, R., {More}, S., {et~al.} 2012, \apj, 749, 38, \dodoi{10.1088/0004-637X/749/1/38}

\bibitem[{{More} {et~al.}(2016){More}, {Verma}, {Marshall}, {More}, {Baeten}, {Wilcox}, {Macmillan}, {Cornen}, {Kapadia}, {Parrish}, {Snyder}, {Davis}, {Gavazzi}, {Lintott}, {Simpson}, {Miller}, {Smith}, {Paget}, {Saha}, {K{\"u}ng}, \& {Collett}}]{more2016a}
{More}, A., {Verma}, A., {Marshall}, P.~J., {et~al.} 2016, \mnras, 455, 1191, \dodoi{10.1093/mnras/stv1965}

\bibitem[{{More} {et~al.}(2024){More}, {Ca{\~n}ameras}, {Jaelani}, {Shu}, {Ishida}, {Wong}, {Inoue}, {Schuldt}, \& {Sonnenfeld}}]{more2024}
{More}, A., {Ca{\~n}ameras}, R., {Jaelani}, A.~T., {et~al.} 2024, \mnras, 533, 525, \dodoi{10.1093/mnras/stae1597}

\bibitem[{{Nordin} {et~al.}(2014){Nordin}, {Rubin}, {Richard}, {Rykoff}, {Aldering}, {Amanullah}, {Atek}, {Barbary}, {Deustua}, {Fakhouri}, {Fruchter}, {Goobar}, {Hook}, {Hsiao}, {Huang}, {Kneib}, {Lidman}, {Meyers}, {Perlmutter}, {Saunders}, {Spadafora}, {Suzuki}, \& {Supernova Cosmology Project}}]{nordin2014a}
{Nordin}, J., {Rubin}, D., {Richard}, J., {et~al.} 2014, \mnras, 440, 2742, \dodoi{10.1093/mnras/stu376}

\bibitem[{{O'Donnell} {et~al.}(2022){O'Donnell}, {Wilkinson}, {Diehl}, {Aros-Bunster}, {Bechtol}, {Birrer}, {Buckley-Geer}, {Carnero Rosell}, {Carrasco Kind}, {da Costa}, {Gonzalez Lozano}, {Gruendl}, {Hilton}, {Lin}, {Lindgren}, {Martin}, {Pieres}, {Rykoff}, {Sevilla-Noarbe}, {Sheldon}, {Sif{\'o}n}, {Tucker}, {Yanny}, {Abbott}, {Aguena}, {Allam}, {Andrade-Oliveira}, {Annis}, {Bertin}, {Brooks}, {Burke}, {Carretero}, {Costanzi}, {De Vicente}, {Desai}, {Dietrich}, {Eckert}, {Everett}, {Ferrero}, {Flaugher}, {Fosalba}, {Frieman}, {Garc{\'\i}a-Bellido}, {Gaztanaga}, {Gerdes}, {Gruen}, {Gschwend}, {Gill}, {Gutierrez}, {Hinton}, {Hollowood}, {Honscheid}, {James}, {Jeltema}, {Kuehn}, {Lahav}, {Lima}, {Maia}, {Marshall}, {Melchior}, {Menanteau}, {Miquel}, {Morgan}, {Nord}, {Ogando}, {Paz-Chinch{\'o}n}, {Pereira}, {Plazas Malag{\'o}n}, {Rodriguez-Monroy}, {Romer}, {Roodman}, {Sanchez}, {Scarpine}, {Schubnell}, {Serrano}, {Smith}, {Suchyta}, {Swanson}, {Tarle}, {Thomas}, {To}, \& {Varga}}]{odonnell2022a}
{O'Donnell}, J.~H., {Wilkinson}, R.~D., {Diehl}, H.~T., {et~al.} 2022, \apjs, 259, 27, \dodoi{10.3847/1538-4365/ac470b}

\bibitem[{{Oguri} \& {Marshall}(2010)}]{oguri2010a}
{Oguri}, M., \& {Marshall}, P.~J. 2010, \mnras, 405, 2579, \dodoi{10.1111/j.1365-2966.2010.16639.x}

\bibitem[{{Pascale} {et~al.}(2024){Pascale}, {Frye}, {Pierel}, {Chen}, {Kelly}, {Cohen}, {Windhorst}, {Riess}, {Kamieneski}, {Diego}, {Meena}, {Cha}, {Oguri}, {Zitrin}, {Jee}, {Foo}, {Leimbach}, {Koekemoer}, {Conselice}, {Dai}, {Goobar}, {Siebert}, {Strolger}, \& {Willner}}]{pascale2024a}
{Pascale}, M., {Frye}, B.~L., {Pierel}, J. D.~R., {et~al.} 2024, arXiv e-prints, arXiv:2403.18902, \dodoi{10.48550/arXiv.2403.18902}

\bibitem[{Pascale {et~al.}(2025)Pascale, Frye, Pierel, Chen, Kelly, Cohen, Windhorst, Riess, Kamieneski, Diego, Meena, Cha, Oguri, Zitrin, Jee, Foo, Leimbach, Koekemoer, Conselice, Dai, Goobar, Siebert, Strolger, \& Willner}]{Pascale_2025}
Pascale, M., Frye, B.~L., Pierel, J. D.~R., {et~al.} 2025, The Astrophysical Journal, 979, 13, \dodoi{10.3847/1538-4357/ad9928}

\bibitem[{{Patel} {et~al.}(2014){Patel}, {McCully}, {Jha}, {Rodney}, {Jones}, {Graur}, {Merten}, {Zitrin}, {Riess}, {Matheson}, {Sako}, {Holoien}, {Postman}, {Coe}, {Bartelmann}, {Balestra}, {Ben{\'\i}tez}, {Bouwens}, {Bradley}, {Broadhurst}, {Cenko}, {Donahue}, {Filippenko}, {Ford}, {Garnavich}, {Grillo}, {Infante}, {Jouvel}, {Kelson}, {Koekemoer}, {Lahav}, {Lemze}, {Maoz}, {Medezinski}, {Melchior}, {Meneghetti}, {Molino}, {Moustakas}, {Moustakas}, {Nonino}, {Rosati}, {Seitz}, {Strolger}, {Umetsu}, \& {Zheng}}]{patel2014a}
{Patel}, B., {McCully}, C., {Jha}, S.~W., {et~al.} 2014, \apj, 786, 9, \dodoi{10.1088/0004-637X/786/1/9}

\bibitem[{{Petrillo} {et~al.}(2019){Petrillo}, {Tortora}, {Vernardos}, {Koopmans}, {Verdoes Kleijn}, {Bilicki}, {Napolitano}, {Chatterjee}, {Covone}, {Dvornik}, {Erben}, {Getman}, {Giblin}, {Heymans}, {de Jong}, {Kuijken}, {Schneider}, {Shan}, {Spiniello}, \& {Wright}}]{petrillo2019a}
{Petrillo}, C.~E., {Tortora}, C., {Vernardos}, G., {et~al.} 2019, \mnras, 484, 3879, \dodoi{10.1093/mnras/stz189}

\bibitem[{{Pierel} \& {Rodney}(2019)}]{pierel2019a}
{Pierel}, J.~D.~R., \& {Rodney}, S. 2019, \apj, 876, 107, \dodoi{10.3847/1538-4357/ab164a}

\bibitem[{{Pierel} {et~al.}(2023){Pierel}, {Arendse}, {Ertl}, {Huang}, {Moustakas}, {Schuldt}, {Shajib}, {Shu}, {Birrer}, {Bronikowski}, {Hjorth}, {Suyu}, {Agarwal}, {Agnello}, {Bolton}, {Chakrabarti}, {Cold}, {Courbin}, {Della Costa}, {Dhawan}, {Engesser}, {Fox}, {Gall}, {Gomez}, {Goobar}, {Jha}, {Jimenez}, {Johansson}, {Larison}, {Li}, {Marques-Chaves}, {Mao}, {Mazzali}, {Perez-Fournon}, {Petrushevska}, {Poidevin}, {Rest}, {Sheu}, {Shirley}, {Silver}, {Storfer}, {Strolger}, {Treu}, {Wojtak}, \& {Zenati}}]{pierel2023a}
{Pierel}, J.~D.~R., {Arendse}, N., {Ertl}, S., {et~al.} 2023, \apj, 948, 115, \dodoi{10.3847/1538-4357/acc7a6}

\bibitem[{{Pierel} {et~al.}(2024){Pierel}, {Newman}, {Dhawan}, {Gu}, {Joshi}, {Li}, {Schuldt}, {Strolger}, {Suyu}, {Caminha}, {Cohen}, {Diego}, {Dsilva}, {Ertl}, {Frye}, {Granata}, {Grillo}, {Koekemoer}, {Li}, {Robotham}, {Summers}, {Treu}, {Windhorst}, {Zitrin}, {Agarwal}, {Agrawal}, {Arendse}, {Belli}, {Burns}, {Ca{\~n}ameras}, {Chakrabarti}, {Chen}, {Collett}, {Coulter}, {Ellis}, {Engesser}, {Foo}, {Fox}, {Gall}, {Garuda}, {Gezari}, {Gomez}, {Glazebrook}, {Hjorth}, {Huang}, {Jha}, {Kamieneski}, {Kelly}, {Larison}, {Moustakas}, {Pascale}, {P{\'e}rez-Fournon}, {Petrushevska}, {Poidevin}, {Rest}, {Shahbandeh}, {Shajib}, {Siebert}, {Storfer}, {Talbot}, {Wang}, {Wevers}, \& {Zenati}}]{pierel2024a}
{Pierel}, J.~D.~R., {Newman}, A.~B., {Dhawan}, S., {et~al.} 2024, arXiv e-prints, arXiv:2404.02139, \dodoi{10.48550/arXiv.2404.02139}

\bibitem[{{Planck Collaboration} {et~al.}(2020){Planck Collaboration}, {Aghanim}, {Akrami}, {Ashdown}, {Aumont}, {Baccigalupi}, {Ballardini}, {Banday}, {Barreiro}, {Bartolo}, {Basak}, {Battye}, {Benabed}, {Bernard}, {Bersanelli}, {Bielewicz}, {Bock}, {Bond}, {Borrill}, {Bouchet}, {Boulanger}, {Bucher}, {Burigana}, {Butler}, {Calabrese}, {Cardoso}, {Carron}, {Challinor}, {Chiang}, {Chluba}, {Colombo}, {Combet}, {Contreras}, {Crill}, {Cuttaia}, {de Bernardis}, {de Zotti}, {Delabrouille}, {Delouis}, {Di Valentino}, {Diego}, {Dor{\'e}}, {Douspis}, {Ducout}, {Dupac}, {Dusini}, {Efstathiou}, {Elsner}, {En{\ss}lin}, {Eriksen}, {Fantaye}, {Farhang}, {Fergusson}, {Fernandez-Cobos}, {Finelli}, {Forastieri}, {Frailis}, {Fraisse}, {Franceschi}, {Frolov}, {Galeotta}, {Galli}, {Ganga}, {G{\'e}nova-Santos}, {Gerbino}, {Ghosh}, {Gonz{\'a}lez-Nuevo}, {G{\'o}rski}, {Gratton}, {Gruppuso}, {Gudmundsson}, {Hamann}, {Handley}, {Hansen}, {Herranz}, {Hildebrandt}, {Hivon}, {Huang}, {Jaffe}, {Jones}, {Karakci}, {Keih{\"a}nen},
  {Keskitalo}, {Kiiveri}, {Kim}, {Kisner}, {Knox}, {Krachmalnicoff}, {Kunz}, {Kurki-Suonio}, {Lagache}, {Lamarre}, {Lasenby}, {Lattanzi}, {Lawrence}, {Le Jeune}, {Lemos}, {Lesgourgues}, {Levrier}, {Lewis}, {Liguori}, {Lilje}, {Lilley}, {Lindholm}, {L{\'o}pez-Caniego}, {Lubin}, {Ma}, {Mac{\'\i}as-P{\'e}rez}, {Maggio}, {Maino}, {Mandolesi}, {Mangilli}, {Marcos-Caballero}, {Maris}, {Martin}, {Martinelli}, {Mart{\'\i}nez-Gonz{\'a}lez}, {Matarrese}, {Mauri}, {McEwen}, {Meinhold}, {Melchiorri}, {Mennella}, {Migliaccio}, {Millea}, {Mitra}, {Miville-Desch{\^e}nes}, {Molinari}, {Montier}, {Morgante}, {Moss}, {Natoli}, {N{\o}rgaard-Nielsen}, {Pagano}, {Paoletti}, {Partridge}, {Patanchon}, {Peiris}, {Perrotta}, {Pettorino}, {Piacentini}, {Polastri}, {Polenta}, {Puget}, {Rachen}, {Reinecke}, {Remazeilles}, {Renzi}, {Rocha}, {Rosset}, {Roudier}, {Rubi{\~n}o-Mart{\'\i}n}, {Ruiz-Granados}, {Salvati}, {Sandri}, {Savelainen}, {Scott}, {Shellard}, {Sirignano}, {Sirri}, {Spencer}, {Sunyaev}, {Suur-Uski}, {Tauber}, {Tavagnacco},
  {Tenti}, {Toffolatti}, {Tomasi}, {Trombetti}, {Valenziano}, {Valiviita}, {Van Tent}, {Vibert}, {Vielva}, {Villa}, {Vittorio}, {Wandelt}, {Wehus}, {White}, {White}, {Zacchei}, \& {Zonca}}]{planck2020}
{Planck Collaboration}, {Aghanim}, N., {Akrami}, Y., {et~al.} 2020, \aap, 641, A6, \dodoi{10.1051/0004-6361/201833910}

\bibitem[{{Refsdal}(1964)}]{refsdal1964a}
{Refsdal}, S. 1964, \mnras, 128, 307, \dodoi{10.1093/mnras/128.4.307}

\bibitem[{{Riess} {et~al.}(2021){Riess}, {Casertano}, {Yuan}, {Bowers}, {Macri}, {Zinn}, \& {Scolnic}}]{riess2021a}
{Riess}, A.~G., {Casertano}, S., {Yuan}, W., {et~al.} 2021, \apjl, 908, L6, \dodoi{10.3847/2041-8213/abdbaf}

\bibitem[{{Rodney} {et~al.}(2016){Rodney}, {Strolger}, {Kelly}, {Brada{\v{c}}}, {Brammer}, {Filippenko}, {Foley}, {Graur}, {Hjorth}, {Jha}, {McCully}, {Molino}, {Riess}, {Schmidt}, {Selsing}, {Sharon}, {Treu}, {Weiner}, \& {Zitrin}}]{rodney2016a}
{Rodney}, S.~A., {Strolger}, L.~G., {Kelly}, P.~L., {et~al.} 2016, \apj, 820, 50, \dodoi{10.3847/0004-637X/820/1/50}

\bibitem[{Rojas {et~al.}(2021)Rojas, Savary, Clément, Maus, Courbin, Lemon, Chan, Vernardos, Joseph, Cañameras, \& Galan}]{rojas2021a}
Rojas, K., Savary, E., Clément, B., {et~al.} 2021, Strong lens systems search in the Dark Energy Survey using Convolutional Neural Networks.
\newblock \doarXiv{2109.00014}

\bibitem[{{Rojas} {et~al.}(2022){Rojas}, {Savary}, {Cl{\'e}ment}, {Maus}, {Courbin}, {Lemon}, {Chan}, {Vernardos}, {Joseph}, {Ca{\~n}ameras}, \& {Galan}}]{rojas2022a}
{Rojas}, K., {Savary}, E., {Cl{\'e}ment}, B., {et~al.} 2022, \aap, 668, A73, \dodoi{10.1051/0004-6361/202142119}

\bibitem[{{Rubin} {et~al.}(2018){Rubin}, {Hayden}, {Huang}, {Aldering}, {Amanullah}, {Barbary}, {Boone}, {Brodwin}, {Deustua}, {Dixon}, {Eisenhardt}, {Fruchter}, {Gonzalez}, {Goobar}, {Gupta}, {Hook}, {Jee}, {Kim}, {Kowalski}, {Lidman}, {Linder}, {Luther}, {Nordin}, {Pain}, {Perlmutter}, {Raha}, {Rigault}, {Ruiz-Lapuente}, {Saunders}, {Sofiatti}, {Spadafora}, {Stanford}, {Stern}, {Suzuki}, {Williams}, \& {Supernova Cosmology Project}}]{rubin2018a}
{Rubin}, D., {Hayden}, B., {Huang}, X., {et~al.} 2018, \apj, 866, 65, \dodoi{10.3847/1538-4357/aad565}

\bibitem[{Sandler {et~al.}(2018)Sandler, Howard, Zhu, Zhmoginov, \& Chen}]{Sandler2018}
Sandler, M., Howard, A.~G., Zhu, M., Zhmoginov, A., \& Chen, L. 2018, CoRR, abs/1801.04381

\bibitem[{{Savary} {et~al.}(2022){Savary}, {Rojas}, {Maus}, {Cl{\'e}ment}, {Courbin}, {Gavazzi}, {Chan}, {Lemon}, {Vernardos}, {Ca{\~n}ameras}, {Schuldt}, {Suyu}, {Cuillandre}, {Fabbro}, {Gwyn}, {Hudson}, {Kilbinger}, {Scott}, \& {Stone}}]{savary2022a}
{Savary}, E., {Rojas}, K., {Maus}, M., {et~al.} 2022, \aap, 666, A1, \dodoi{10.1051/0004-6361/202142505}

\bibitem[{{Sharma} \& {Linder}(2022)}]{sharma2022a}
{Sharma}, D., \& {Linder}, E.~V. 2022, arXiv e-prints, arXiv:2204.03020.
\newblock \doarXiv{2204.03020}

\bibitem[{{Sheu} {et~al.}(2024){Sheu}, {Huang}, {Cikota}, {Suzuki}, {Palmese}, {Schlegel}, \& {Storfer}}]{sheu2024}
{Sheu}, W., {Huang}, X., {Cikota}, A., {et~al.} 2024, \apj, 973, 24, \dodoi{10.3847/1538-4357/ad5dad}

\bibitem[{{Sheu} {et~al.}(2023){Sheu}, {Huang}, {Cikota}, {Suzuki}, {Schlegel}, \& {Storfer}}]{sheu2023a}
---. 2023, arXiv e-prints, arXiv:2301.03578, \dodoi{10.48550/arXiv.2301.03578}

\bibitem[{{Shu} {et~al.}(2022){Shu}, {Ca{\~n}ameras}, {Schuldt}, {Suyu}, {Taubenberger}, {Inoue}, \& {Jaelani}}]{shu2022a}
{Shu}, Y., {Ca{\~n}ameras}, R., {Schuldt}, S., {et~al.} 2022, \aap, 662, A4, \dodoi{10.1051/0004-6361/202243203}

\bibitem[{{Shu} {et~al.}(2017){Shu}, {Brownstein}, {Bolton}, {Koopmans}, {Treu}, {Montero-Dorta}, {Auger}, {Czoske}, {Gavazzi}, {Marshall}, \& {Moustakas}}]{shu2017a}
{Shu}, Y., {Brownstein}, J.~R., {Bolton}, A.~S., {et~al.} 2017, \apj, 851, 48, \dodoi{10.3847/1538-4357/aa9794}

\bibitem[{{Sonnenfeld} {et~al.}(2019){Sonnenfeld}, {Wang}, \& {Bahcall}}]{sonnenfeld2019a}
{Sonnenfeld}, A., {Wang}, W., \& {Bahcall}, N. 2019, \aap, 622, A30, \dodoi{10.1051/0004-6361/201834260}

\bibitem[{{Sonnenfeld} {et~al.}(2018){Sonnenfeld}, {Chan}, {Shu}, {More}, {Oguri}, {Suyu}, {Wong}, {Lee}, {Coupon}, {Yonehara}, {Bolton}, {Jaelani}, {Tanaka}, {Miyazaki}, \& {Komiyama}}]{sonnenfeld2018a}
{Sonnenfeld}, A., {Chan}, J.~H.~H., {Shu}, Y., {et~al.} 2018, \pasj, 70, S29, \dodoi{10.1093/pasj/psx062}

\bibitem[{{Sonnenfeld} {et~al.}(2020){Sonnenfeld}, {Verma}, {More}, {Allen}, {Baeten}, {Chan}, {Hutchings}, {Jaelani}, {Lee}, {Macmillan}, {Marshall}, {O' Donnell}, {Oguri}, {Rusu}, {Veldthuis}, {Wong}, {Cornen}, {Davis}, {McMaster}, {Trouille}, {Lintott}, \& {Miller}}]{sonnenfeld2020a}
{Sonnenfeld}, A., {Verma}, A., {More}, A., {et~al.} 2020, arXiv e-prints, arXiv:2004.00634.
\newblock \doarXiv{2004.00634}

\bibitem[{{Stein} {et~al.}(2022){Stein}, {Blaum}, {Harrington}, {Medan}, \& {Luki{\'c}}}]{stein2021a}
{Stein}, G., {Blaum}, J., {Harrington}, P., {Medan}, T., \& {Luki{\'c}}, Z. 2022, \apj, 932, 107, \dodoi{10.3847/1538-4357/ac6d63}

\bibitem[{{Storfer} {et~al.}(2024){Storfer}, {Huang}, {Gu}, {Sheu}, {Banka}, {Dey}, {Inchausti Reyes}, {Jain}, {Kwon}, {Lang}, {Lee}, {Meisner}, {Moustakas}, {Myers}, {Tabares-Tarquinio}, {Schlafly}, \& {Schlegel}}]{storfer2024a}
{Storfer}, C., {Huang}, X., {Gu}, A., {et~al.} 2024, \apjs, 274, 16, \dodoi{10.3847/1538-4365/ad527e}

\bibitem[{{Storfer} {et~al.}(2025){Storfer}, {Magnier}, {Huang}, {Rubin}, {Schlegel}, {Banka}, {Chambers}, {Cuillandre}, {de Boer}, {Gavazzi}, {Gwyn}, {Hudson}, {Paek}, \& {Scott}}]{storfer2025}
{Storfer}, C.~J., {Magnier}, E.~A., {Huang}, X., {et~al.} 2025, arXiv e-prints, arXiv:2505.05032, \dodoi{10.48550/arXiv.2505.05032}

\bibitem[{{Suyu} {et~al.}(2020){Suyu}, {Huber}, {Ca{\~n}ameras}, {Schuldt}, {Taubenberger}, {Y{\i}ld{\i}r{\i}m}, {Bonvin}, {Chan}, {Courbin}, {Kromer}, {N{\"o}bauer}, {Sim}, \& {Sluse}}]{suyu2020a}
{Suyu}, S.~H., {Huber}, S., {Ca{\~n}ameras}, R., {et~al.} 2020, arXiv e-prints, arXiv:2002.08378.
\newblock \doarXiv{2002.08378}

\bibitem[{{Szegedy} {et~al.}(2014){Szegedy}, {Liu}, {Jia}, {Sermanet}, {Reed}, {Anguelov}, {Erhan}, {Vanhoucke}, \& {Rabinovich}}]{szegedy2014a}
{Szegedy}, C., {Liu}, W., {Jia}, Y., {et~al.} 2014, arXiv e-prints, arXiv:1409.4842.
\newblock \doarXiv{1409.4842}

\bibitem[{{Talbot} {et~al.}(2021){Talbot}, {Brownstein}, {Dawson}, {Kneib}, \& {Bautista}}]{talbot2021a}
{Talbot}, M.~S., {Brownstein}, J.~R., {Dawson}, K.~S., {Kneib}, J.-P., \& {Bautista}, J. 2021, \mnras, 502, 4617, \dodoi{10.1093/mnras/stab267}

\bibitem[{Tan \& Le(2019)}]{Tan2019}
Tan, M., \& Le, Q.~V. 2019, ArXiv, abs/1905.11946

\bibitem[{{Treu}(2010)}]{treu2010a}
{Treu}, T. 2010, \araa, 48, 87, \dodoi{10.1146/annurev-astro-081309-130924}

\bibitem[{{Vegetti} {et~al.}(2010){Vegetti}, {Czoske}, \& {Koopmans}}]{vegetti2010a}
{Vegetti}, S., {Czoske}, O., \& {Koopmans}, L. V.~E. 2010, \mnras, 407, 225, \dodoi{10.1111/j.1365-2966.2010.16952.x}

\bibitem[{{Vegetti} \& {Koopmans}(2009)}]{vegetti2009a}
{Vegetti}, S., \& {Koopmans}, L.~V.~E. 2009, \mnras, 392, 945, \dodoi{10.1111/j.1365-2966.2008.14005.x}

\bibitem[{{Wagner-Carena} {et~al.}(2022){Wagner-Carena}, {Aalbers}, {Birrer}, {Nadler}, {Darragh-Ford}, {Marshall}, \& {Wechsler}}]{wagner-carena2022}
{Wagner-Carena}, S., {Aalbers}, J., {Birrer}, S., {et~al.} 2022, arXiv e-prints, arXiv:2203.00690.
\newblock \doarXiv{2203.00690}

\bibitem[{{Williams} {et~al.}(2004){Williams}, {Olszewski}, {Lesser}, \& {Burge}}]{williams2004a}
{Williams}, G.~G., {Olszewski}, E., {Lesser}, M.~P., \& {Burge}, J.~H. 2004, in \procspie, Vol. 5492, Ground-based Instrumentation for Astronomy, ed. A.~F.~M. {Moorwood} \& M.~{Iye}, 787--798

\bibitem[{{Wong} {et~al.}(2022){Wong}, {Chan}, {Chao}, {Jaelani}, {Kayo}, {Lee}, {More}, \& {Oguri}}]{wong2022a}
{Wong}, K.~C., {Chan}, J. H.~H., {Chao}, D. C.~Y., {et~al.} 2022, \pasj, 74, 1209, \dodoi{10.1093/pasj/psac065}

\bibitem[{{Wong} {et~al.}(2018){Wong}, {Sonnenfeld}, {Chan}, {Rusu}, {Tanaka}, {Jaelani}, {Lee}, {More}, {Oguri}, {Suyu}, \& {Komiyama}}]{wong2018a}
{Wong}, K.~C., {Sonnenfeld}, A., {Chan}, J. H.~H., {et~al.} 2018, \apj, 867, 107, \dodoi{10.3847/1538-4357/aae381}

\bibitem[{{Wong} {et~al.}(2019){Wong}, {Suyu}, {Chen}, {Rusu}, {Millon}, {Sluse}, {Bonvin}, {Fassnacht}, {Taubenberger}, {Auger}, {Birrer}, {Chan}, {Courbin}, {Hilbert}, {Tihhonova}, {Treu}, {Agnello}, {Ding}, {Jee}, {Komatsu}, {Shajib}, {Sonnenfeld}, {Bland ford}, {Koopmans}, {Marshall}, \& {Meylan}}]{wong2019a}
{Wong}, K.~C., {Suyu}, S.~H., {Chen}, G. C.~F., {et~al.} 2019, arXiv e-prints, arXiv:1907.04869.
\newblock \doarXiv{1907.04869}

\bibitem[{{Yahalomi} {et~al.}(2017){Yahalomi}, {Schechter}, \& {Wambsganss}}]{yahalomi2017a}
{Yahalomi}, D.~A., {Schechter}, P.~L., \& {Wambsganss}, J. 2017, arXiv e-prints, arXiv:1711.07919.
\newblock \doarXiv{1711.07919}

\bibitem[{Zaborowski {et~al.}(2023)Zaborowski, Drlica-Wagner, Ashmead, Wu, Morgan, Bom, Shajib, Birrer, Cerny, Buckley-Geer, Mutlu-Pakdil, Ferguson, Glazebrook, Lozano, Gordon, Martinez, Manwadkar, O'Donnell, Poh, Riley, Sakowska, Santana-Silva, Santiago, Sluse, Tan, Tollerud, Verma, Carballo-Bello, Choi, James, Kuropatkin, Mart{\'\i}nez-V{\'a}zquez, Nidever, Castellon, No{\"e}l, Olsen, Pace, Mau, Yanny, Zenteno, Abbott, Aguena, Alves, Andrade-Oliveira, Bocquet, Brooks, Burke, Carnero~Rosell, Carrasco~Kind, Carretero, Castander, Conselice, Costanzi, Pereira, De~Vicente, Desai, Dietrich, Doel, Everett, Ferrero, Flaugher, Friedel, Frieman, Garc{\'\i}a-Bellido, Gruen, Gruendl, Gutierrez, Hinton, Hollowood, Honscheid, Kuehn, Lin, Marshall, Melchior, Mena-Fern{\'a}ndez, Menanteau, Miquel, Palmese, Paz-Chinch{\'o}n, Pieres, Malag{\'o}n, Prat, Rodriguez-Monroy, Romer, Sanchez, Scarpine, Sevilla-Noarbe, Smith, Suchyta, To, Weaverdyck, \& Collaborations)}]{Zaborowski_2023}
Zaborowski, E.~A., Drlica-Wagner, A., Ashmead, F., {et~al.} 2023, The Astrophysical Journal, 954, 68, \dodoi{10.3847/1538-4357/ace4ba}

\bibitem[{{Zenteno} {et~al.}(2025){Zenteno}, {Kluge}, {Kharkrang}, {Hernandez-Lang}, {Damke}, {Saro}, {Monteiro-Oliveira}, {Carrasco}, {Salvato}, {Comparat}, {Fabricius}, {Snigula}, {Arevalo}, {Cuevas}, {Nilo Castellon}, {Ramirez}, {V{\'e}liz Astudillo}, {Landriau}, {Myers}, {Schlafly}, {Valdes}, {Weaver}, {Mohr}, {Grandis}, {Klein}, {Liu}, {Bulbul}, {Zhang}, {Sanders}, {Bahar}, {Ghirardini}, {Ramos}, \& {Balzer}}]{Zenteno2025}
{Zenteno}, A., {Kluge}, M., {Kharkrang}, R., {et~al.} 2025, \aap, 698, A171, \dodoi{10.1051/0004-6361/202452440}

\bibitem[{{Zhou} {et~al.}(2021){Zhou}, {Newman}, {Mao}, {Meisner}, {Moustakas}, {Myers}, {Prakash}, {Zentner}, {Brooks}, {Duan}, {Landriau}, {Levi}, {Prada}, \& {Tarle}}]{zhou2021a}
{Zhou}, R., {Newman}, J.~A., {Mao}, Y.-Y., {et~al.} 2021, \mnras, 501, 3309, \dodoi{10.1093/mnras/staa3764}

\bibitem[{{Zitrin} {et~al.}(2012){Zitrin}, {Rephaeli}, {Sadeh}, {Medezinski}, {Umetsu}, {Sayers}, {Nonino}, {Morandi}, {Molino}, {Czakon}, \& {Golwala}}]{zitrin2012}
{Zitrin}, A., {Rephaeli}, Y., {Sadeh}, S., {et~al.} 2012, \mnras, 420, 1621, \dodoi{10.1111/j.1365-2966.2011.20155.x}

\end{thebibliography}
\end{document}